\documentclass[aps, prd, superscriptaddress,preprintnumbers,nofootinbib]{revtex4-1}
\usepackage[english]{babel}
\usepackage[T1]{fontenc}
\usepackage[utf8]{inputenc}
\usepackage{verbatim}
\usepackage{empheq}
\usepackage{amssymb}
\usepackage{amsmath}
\usepackage{amsfonts}
\usepackage{float}

\usepackage[colorlinks]{hyperref}
\hypersetup{citecolor=blue,linkcolor=blue,urlcolor=blue}

\usepackage{dcolumn}
\usepackage{graphicx}
\usepackage{etoolbox}
\usepackage{booktabs}
\usepackage{babel}

\usepackage{xcolor}

\newcommand{\bea}{\begin{eqnarray}}
\newcommand{\eea}{\end{eqnarray}}
\newcommand{\nn}{\nonumber \\}

\def\beq{\begin{equation}}
\def\eeq{\end{equation}}

\newcommand{\nc}{\newcommand}

\newcommand{\C}{\mathcal{C}}
\newcommand{\Op}{{\cal O}}
\nc{\vp}{\phi}
\nc{\tvp}{\widetilde{\phi}}
\nc{\vpj }{\mbox{${\vp^\dag i\,\raisebox{2mm}{\boldmath ${}^\leftrightarrow$}\hspace{-4mm} D_\mu\,\vp}$}}
\nc{\vpjt}{\mbox{${\vp^\dag i\,\raisebox{2mm}{\boldmath ${}^\leftrightarrow$}\hspace{-4mm} D_\mu^{\,a}\,\vp}$}}

\def\nn{\nonumber}
\def\gev{\rm GeV}

\begin{document} 

\global\long\def\order#1{\mathcal{O}{\left(#1\right)}}
\global\long\def\d{\mathrm{d}}
\global\long\def\P{P}
\global\long\def\amp{{\mathcal M}}

\def\BNL{High Energy Theory Group, Department of Physics, Brookhaven 
	National Laboratory, Upton, NY 11973, USA }

\title{The Importance of Flavor in SMEFT Electroweak Precision Fits} 

\author{Luigi Bellafronte}
\email[Electronic address: ]{lui.bellafronte@usc.es}
\affiliation{Instituto Galego de F\'{i}sica de Altas Enerx\'{i}as, Universidade de Santiago de Compostela,\\
15782 Santiago de Compostela, Galicia, Spain}

\author{Sally~Dawson}
\email[Electronic address: ]{dawson@bnl.gov}
\affiliation{\BNL}

\author{Pier Paolo Giardino}
\email[Electronic address: ]{pierpaolo.giardino@usc.es }
\affiliation{Instituto Galego de F\'{i}sica de Altas Enerx\'{i}as, Universidade de Santiago de Compostela,\\
15782 Santiago de Compostela, Galicia, Spain}

\begin{abstract}
Effective field theory tools are essential for exploring non-Standard Model physics at the LHC in the absence of the discovery of new light 
particles.  Predictions for observables are typically made at the lowest order in the QCD and electroweak expansions in the Standard Model effective field theory (SMEFT) and often ignore the effects of flavor.  Here, we present results for electroweak precision observables (EWPOs)
at the next-to-leading order QCD and electroweak expansions (NLO)  of the SMEFT with an arbitrary flavor structure for the  fermion operators. Numerical NLO SMEFT fits to 
EWPOs have a strong dependence on the assumed flavor structures and we demonstrate this using various popular assumptions for flavor symmetries.  
\end{abstract}

\maketitle
\newpage

\section{Introduction}

The high luminosity (HL) LHC program will greatly improve our knowledge of the allowed possibilities for physics beyond the Standard Model (SM).  In the absence of the discovery of a new particle, the Standard Model Effective Field Theory (SMEFT) is the tool of choice for describing new physics effects and provides a point of comparison with Standard Model (SM) predictions\cite{Brivio:2017vri,Dedes:2017zog}. The SMEFT is constructed as an expansion in higher dimension operators, with the $SU(3)\times SU(2)\times U(1)$ gauge symmetry unbroken,
\begin{equation}
{\cal L}={\cal L}_{SM}+\Sigma_{k=5}^{\infty}\Sigma_{a=1}^n {{C}_a^k\over \Lambda^{k-4}} \Op_a^k\, .
\label{eq:lsmeft}
\end{equation}
The
 operators of mass-dimension $k$, $\Op_a^k$,  are constructed from SM fields and all of  the effects of the beyond the
 SM (BSM)
  physics  reside in the coefficient functions, $C_a^k$. In a UV complete model, these coefficients will be predicted
  at the high scale $\Lambda$ \cite{deBlas:2017xtg,Brehmer:2015rna,Dawson:2020oco,Corbett:2021eux}. The comparison between the SMEFT and the SM predictions thus
requires SM predictions to the highest possible accuracy, along with SMEFT calculations beyond the leading order (LO). LO calculations  in the SMEFT are automated\cite{Brivio:2020onw,Brivio:2017btx} as are NLO QCD SMEFT calculations \cite{Degrande:2020evl}, but electroweak corrections must be  performed on a case by case basis.

There exist numerous global fits comparing SMEFT predictions with electroweak precision, di-boson, Higgs, and top quark observables\cite{Ethier:2021bye,Ellis:2020unq,Biekoetter:2018ypq,Bagnaschi:2022whn,Ellis:2018gqa}.   However, given the large number of SMEFT operators, it is impossible to do a fully general fit. The SMEFT theory predictions thus encode various assumptions, including the termination of the SMEFT expansion at some order 
in $1/\Lambda^2$, the loop order of the QCD and electroweak expansions, and the flavor structure of fermion operators, among others.  Understanding the uncertainties inherent in these assumptions is crucial for interpreting SMEFT
fits\cite{Brivio:2022pyi}.

The assumptions about the flavor structure introduce significant model dependence into  the SMEFT predictions.  It
is straightforward to implement a general flavor structure for the tree level predictions for observables\cite{Efrati:2015eaa,Falkowski:2019hvp,Greljo:2022cah,Faroughy:2020ina,Bruggisser:2021duo,Crivellin:2020ebi,Crivellin:2022rhw} and the one-loop ${\overline{MS}}$ renormalization of the dimension-6 coefficient functions is known for an 
arbitrary flavor structure\cite{Jenkins:2013zja,Jenkins:2013wua,Alonso:2013hga}. 
The one-loop 
next-to-leading order  (NLO) electroweak predictions for physical observables, however, typically involve a large number of 4-fermion and 2-fermion operators with potentially complicated flavor structures.  In addition, the fermion operators introduce new subtleties in the renormalization procedure. Existing calculations of the one-loop electroweak corrections to EWPO, Higgs, and di-boson data do not include the most general flavor structure for the 2- quark and 4- quark operators in the loops\cite{Dawson:2018pyl,Dawson:2018liq,Cullen:2020zof,Dedes:2019bew,Dedes:2018seb,Hartmann:2016pil,Hartmann:2015aia}.  Here, we generalize our previous NLO SMEFT calculation of EWPOs\cite{Dawson:2019clf,Dawson:2018jlg,Dawson:2022bxd} which included 4-fermion operators, but not 2-fermion operators, to allow for an arbitrary flavor structure.
  The corrections to the EWPOs from 4-fermion operators in the $U(3)^5$ symmetric case are in \cite{Hartmann:2016pil}.
The role of flavor assumptions in fits to top and bottom quark data has been extensively examined in the literature and those analyses are complementary to that
presented here\cite{Brivio:2019ius,Alasfar:2020mne,Boughezal:2019xpp,Dawson:2022bxd}. 

 We set the CKM matrix to be diagonal, which implies that only operators containing pairs
of identical flavor fermions contribute.  We further work to linear order in the SMEFT coefficients and set all masses other than the top quark to be 0. We want to remark that these two conditions constrain the light fermion Yukawa couplings to be zero, without additional assumptions, since the SMEFT operators that would induce a modification to the SM Yukawas do not interfere with the SM amplitudes at linear order.

In Section \ref{sec:smeft}, we review the SMEFT framework and the role of flavor assumptions.  Here we construct the general form of  the 2-fermion and 4-fermion coefficients that occur in the one-loop computations of EWPOs  corresponding to different flavor anzatz . We describe the NLO electroweak and QCD renormalization of EWPOs in detail in Section \ref{sec:nlo} with emphasis on the roles of the 2- and 4-fermion operators.  Section \ref{sec:results} contains phenomenology results and demonstrates the important role of flavor assumptions in the SMEFT fits\cite{Bruggisser:2022rhb}. We see that even including more general flavor assumptions than earlier analyses, many blind directions remain in the fits.  Finally, Section \ref{sec:conc} contains some conclusions.  Numerical results for the EWPOs, along with $H\rightarrow Z\gamma$ and $H\rightarrow \gamma\gamma$, with arbitrary flavor structures at NLO in the SMEFT are contained in supplemental material.
\section{SMEFT}
\label{sec:smeft}
Our calculations are done using the dimension-6 SMEFT Lagrangian and the Warsaw basis\cite{Grzadkowski:2010es}.
  Retaining only the dimension-6 operators (and dropping the superscript $k$),
   we calculate  observables, $O_b$, to one-loop as an expansion in
  ${1\over \Lambda^2}$ and keep only the linear terms since the SMEFT is renormalizable order by order in powers of 
  ${1\over \Lambda^2}$,
  \begin{equation}
  O_b=O_{b,SM}+\Sigma_{a=1}^n {{C}_a\over \Lambda^{2}} \beta_{ab} ,\,
  \end{equation}
  where $\beta_{ab}$ is process dependent and depends on the kinematic invariants and the input parameters,
  and $O_{b,SM}$ is the SM prediction.
The dimension-6 SMEFT framework is described in detail in the literature, and  here we review only those aspects
relevant to the current calculations.  We use the Feynman rules from Ref. \cite{Dedes:2017zog}, with general flavor structures for the 2- and 4-fermion operators,
although we assume that the CKM matrix is diagonal, which has implications for the fermion structures of the operators that contribute to our calculation, as we will see.  Furthermore, we assume that the SMEFT does not introduce new sources of CP violation, that is we assume the coefficients of all the operators to be real.

Our goal is
the dimension-6 SMEFT calculation of NLO QCD and NLO electroweak corrections to electroweak precision observables (EWPOs)
with arbitrary flavor structures for the operators involving fermions.  The technical details are in the next section and in the following sub-sections we discuss the effects of flavor on the predictions.  

In the following, fermion fields are written using the notation
 \begin{eqnarray}
&(q^{i})^T=({\overline {u}}_L^i, {\overline {d}}_L^i) \qquad &u^i=u_R, c_R, t_R\nonumber  \\
&(l^{i})^T=({\overline {\nu}}_L^i, {\overline {e}}_L^i) \qquad &d^i=d_R, s_R, b_R \nonumber \\
&\qquad&e^i=e_R, \mu_R, \tau_R\, , 
\end{eqnarray}
with $i=1,2,3$ the generation (flavor) index. Moreover, we define the Higgs doublet as 
\bea
\phi=\left(\begin{array}{c}\varphi^+\\ \frac{1}{\sqrt{2}}(v+h+i\varphi^0)\end{array}\right),
\eea
where $h$, $\varphi^+$ and $\varphi^0$ are the Higgs and Goldstone bosons, and $v$ is the electroweak vacuum expectation value.

\subsection{EWPO}
In complete generality, the Warsaw basis contains 2499 baryon number conserving dimension-6 operators.  Much of the proliferation of operators is associated with the flavor structure.  Of course, most of the flavor structures will not contribute to a given observable and we begin by considering EWPOs at tree level.   The $Z$ and $W$ boson pole observables that we consider are,
\begin{eqnarray}
&&M_W, \Gamma_W, \Gamma_Z,  \sigma_h,  R_e,  R_\mu, R_\tau, R_s, R_c, R_b,  A_e, A_\mu, A_\tau, A_s, A_c, A_b, A_{e,FB},A_{\mu,FB},A_{\tau,FB}, A_{FB,s}, A_{FB,c},A_{FB,b}\, .
\label{eq:quan}
\end{eqnarray} 
Note that we do not use the effective mixing angle in our fits, since it is derived from the asymmetries.

The operators that contribute to the EWPOs at LO are comprised of two bosonic operators with no flavor structure ($O_{\phi W B}$ and $O_{\phi D}$) and 8 fermionic operators, of which there are 7 operators with 2 fermionic indices (the 2-fermion operators), and 1 operator with 4 fermionic indices (the 4-fermion operator). The explicit forms of the operators that appear at LO are reported in Table \ref{tab:opdef}.  These operators change the couplings of the $Z$ and $W$ bosons to fermions, and explicit predictions for the measured quantities of Eq. \ref{eq:quan} are given in  Appendix A of Ref. \cite{Dawson:2018liq}. 
The only 4-fermion operator that is relevant for EWPOs at LO is ${\Op_{ll}}[ijkl]$ which has the symmetry
${\Op_{ll}}[ijkl]={\Op_{ll}}[klij]$ and only ${\Op_{ll}}[2112]={\Op_{ll}}[1221]$ contributes to the observables of Eq.\ref{eq:quan}.  The indices ($i,j,k,l=1,2,3$) refer to the fermion generation.)
Dimension-6  operators involving the electron and the  muon  give contributions to the decay of the $\mu$, changing the relation between the 
vev, $v$, and the Fermi constant $G_\mu$, 
\begin{eqnarray}
G_\mu
\equiv {1\over \sqrt{2} v^2}-\frac1 {\sqrt{2} 
\Lambda^2}\C_{ll}[1221]+
{1\over \sqrt{2} \Lambda^2}\biggl(\C_{\phi l}^{(3)}[11]+\C_{\phi l}^{(3)}[22]\biggr)\, .
\label{eq:gdef}
\end{eqnarray}

\begin{table}[t] 
\centering
\renewcommand{\arraystretch}{1.5}
\begin{tabular}{||c|c||c|c||c|c||} 
\hline \hline
${\Op_{ll}}[ijkl]$                   & $(\bar l_i\gamma_\mu l_j)(\bar l_k \gamma^\mu l_l)$  &    
  ${\Op}_{\phi W B}$ 
 &$ (\vp^\dag \tau^a \vp)\, W^a_{\mu\nu} B^{\mu\nu}$  &
$\Op_{\vp D}$   & $\left(\vp^\dag D^\mu\vp\right)^* \left(\vp^\dag D_\mu\vp\right)$ 
\\
\hline 
   ${\Op}_{\phi e}[ij]$  &   $(\vpj) (\overline {e}_{Ri}\gamma^\mu e_{Rj})$  & ${\Op}_{\phi u}[ij]$ & $(\vpj) (\overline {u}_{R i}\gamma^\mu u_{Rj })$ &
  ${\Op}_{\phi d}[ij]$
       & $(\vpj) (\overline {d}_{Ri }\gamma^\mu d_{Rj})$
  \\ \hline 
             ${\Op}_{\phi q}^{(3)}[ij]$ & $(\vpjt)(\bar q_i \tau^a \gamma^\mu q_j)$  &${\Op}_{\phi q}^{(1)}[ij]$
      &$(\vpj)(\bar q_i \tau^a \gamma^\mu q_j)$  &
 ${\Op_{\vp l}^{(3)}}[ij]$      & $(\vpjt)(\bar l_i \tau^a \gamma^\mu l_j)$
 \\
\hline
${\Op}_{\phi l}^{(1)}[ij]$
      &  $(\vpj)(\bar l_i \tau^a \gamma^\mu l_j)$ &
    && &

\\
\hline \hline
\end{tabular}
\caption{Dimension-6 operators  contributing to the  $Z$ and $W$ pole observables of this study at tree level. $i,j,k,l =1,2,3$ are generation indices. We define $\vpj=i\phi^\dagger(D_{\mu} \phi)
-i(D_{\mu}\phi)^\dagger \phi$
and
$\vpjt=i\phi^\dagger \tau ^a D_{\mu} \phi
-i(D_{\mu}\phi)^\dagger \tau^a\phi$.  \label{tab:opdef}}
\end{table}

At NLO, the EWPOs of Eq. \ref{eq:quan} receive contributions from 22 additional operators (which are defined in Ref. \cite{Dawson:2018liq}), which we classify according to the number of fermions:
\begin{itemize}
\item 4 bosonic operators:
\begin{equation}{\Op}_{\phi B},\,{\Op}_{\phi W}
,\,{\Op}_{\square}\,,{\Op}_{W}\, .
\end{equation}
\item 2 2-fermion operators:
\begin{equation}
{\Op}_{uB}[ij],\,{\Op}_{uW}[ij]\,.
\end{equation}
Notice that only ${\Op}_{uB}[33]$ and ${\Op}_{uW}[33]$ contribute to the EWPOs at NLO if all fermions except the top are massless.
\item 16 4-fermion operators: 
\begin{eqnarray}&&
{\Op}_{ed}[ijkl],\,{\Op}_{ee}[ijkl],\,{\Op}_{eu}[ijkl],\,{\Op}_{lu}[ijkl]\,,{\Op}_{ld}[ijkl],\,{\Op}_{le}[ijkl],\,{\Op}_{lq}^{(1)}[ijkl],\,{\Op}_{lq}^{(3)}[ijkl]\nonumber \\
&&{\Op}_{qe}[ijkl]\,  , {\Op}_{qd}^{(1)}[ijkl]\,, {\Op}_{qq}^{(3)}[ijkl]\,,{\Op}_{qq}^{(1)}[ijkl]\,,{\Op}_{qu}^{(1)}[ijkl]\,,
{\Op}_{ud}^{(1)}[ijkl]\,
,{\Op}_{uu}[ijkl]\,\,, {\Op}_{dd}[ijkl] \, .
\end{eqnarray}
Five of the NLO-generated 4-fermion operators have a flavor symmetry,
\begin{equation}
{\Op}_{ee}[ijkl],\, {\Op}_{qq}^{(3)}[ijkl],\,{\Op}_{qq}^{(1)}[ijkl]\,
,{\Op}_{uu}[ijkl],\, {\Op}_{dd}[ijkl]\equiv   {\Op}_Y[ijkl]={\Op}_Y[klij] .
\label{eq:fsym}
\end{equation}

\end{itemize}

It is convenient to categorize the operators according to their dependence on the flavor since only specific structures contribute to the EWPOs at NLO:
\begin{itemize}
\item
A) 2-fermion operators: ${\Op}_X[ij]\equiv{\Op}_{\phi e}[ij],\, {\Op}_{\phi u}[ij],\, {\Op}_{\phi d}[ij],\, 
 {\Op}_{\phi q}^{(3)}[ij],\,
  {\Op}_{\phi q}^{(1)}[ij],\, {\Op}_{\phi l}^{(3)}[ij] ,\, {\Op}_{\phi l}^{(1)}[ij],\,{\Op}_{uB}[ij],\,{\Op}_{uW}[ij]\, .$
We consider the  CKM matrix to be diagonal which has the consequence that
 the coefficients of the operators in Class A have diagonal flavor structures:
\bea
{\C}_X[ij]=E^{(i)}_X\delta_{ij} ,\,\quad i,j=1,2,3,
\eea 
resulting in 3 independent coefficients for each operator. Notice that the operator ${\Op}_{\phi u}[33]$ first contributes to the EWPOs  at NLO. Furthermore, as noted above, ${\Op}_{uB}[ij]$ and ${\Op}_{uW}[ij]$ enter in our calculations only with coefficients ${\C}_{uB}[33]$ and ${\C}_{uW}[33]$ respectively. Together there are 23 independent coefficients in Class A. 
 \item
 B) 4-fermion operators involving only identical fermion representations: 
 ${\Op}_Y[ijkl]\equiv$ ${\Op}_{ee}[ijkl]$, ${\Op}_{qq}^{(3)}[ijkl]$, ${\Op}_{qq}^{(1)}[ijkl]$, ${\Op}_{uu}[ijkl]$,  ${\Op}_{dd}[ijkl]$, ${\Op}_{ll}[ijkl]$. Since they stem from the combination of two fermion currents belonging to identical representations once we require that  the CKM matrix is the unit matrix, there are only two ways flavor is allowed to "flow" through these operators:  ${\Op}_Y[iijj]$ and ${\Op}_Y[ijji]$. Furthermore, these operators are subject to the flavor symmetry in Eq. \ref{eq:fsym}, resulting in the coefficients having the flavor structure,
\bea
&&{\C}_Y[iiii]=F_Y^{(i)},\,{\C}_Y[iijj]=A_Y^{(ij)},\,{\C}_Y[ijji]=B_Y^{(ij)},\nn\\
&&A_Y^{(ji)}=A_Y^{(ij)},\,B_Y^{(ji)}=B_Y^{(ij)},\, \quad  i\neq j\, \&\, i,j=1,2,3,
\eea
resulting in 9 independent coefficients for each operator. However, the flavor structure of ${\Op}_{ee}[ijkl]$ is further constrained by the Fiertz identity  $ ({\overline{e}}_i\gamma^\mu e_j)({\overline{e}}_k\gamma^\mu e_l)= ({\overline{e}}_i\gamma^\mu e_l)({\overline{e}}_k\gamma^\mu e_j)$, which imposes the equality $A^{(ij)}_{ee}=B^{(ij)}_{ee}$, reducing the number of independent coefficients for ${\Op}_{ee}[ijkl]$ to 6. The only coefficient of Class B that does not contribute to the EWPOs at NLO is ${\C}_{uu}[3333]$. In total, we have 50 independent coefficients in Class B contributing. 
 \item
C) 4-fermion operators with 2 different fermion representations:
${\Op}_Z[ijkl]\equiv$ ${\Op}_{ed}[ijkl]$, ${\Op}_{eu}[ijkl]$, ${\Op}_{lu}[ijkl]$, ${\Op}_{ld}[ijkl]$, ${\Op}_{le}[ijkl]$, 
${\Op}_{lq}^{(1)}[ijkl]$, ${\Op}_{lq}^{(3)}[ijkl]$, ${\Op}_{qe}[ijkl]$, ${\Op}_{qd}^{(1)}[ijkl]$, ${\Op}_{qu}^{(1)}[ijkl]$,
 ${\Op}_{ud}^{(1)}[ijkl]$.
 For these operators, our choice of a diagonal CKM matrix requires that the flavor must flow only in one way: ${\Op}_Y[iijj]$. Therefore, the coefficients of these operators have the flavor structure:
 \beq
 {\C}_Z[iijj]=D_Z^{(ij)},\,\quad  i,j=1,2,3,
 \eeq
with no further restrictions, corresponding to 9 independent coefficients for each operator, for a total of 99 independent coefficients in Class C that contribute to the observables of Eq. \ref{eq:quan} at NLO. \end{itemize}

In the most general flavor case, we see that EWPOs computed to NLO receive contributions from 178 independent coefficients: 6 from bosonic operators, 23 from 2-fermion operators, and 149 from 4-fermion operators. We next consider different flavor assumptions for the fermionic operators in order to reduce the number of operators that need to be considered.  The flavor assumptions we consider are $U(3)^5$, minimal flavor violation (MFV), $U(2)^5$, third generation centric, third generation phobic, third generation phobic + $U(2)^5$, and a flavorless structure.  We will discuss each of these assumptions in detail in the following sub-sections. 

\subsection{Flavor Assumptions: $U(3)^5$}
\label{sec:g3}
 In the absence of Yukawa couplings, the SM fermions have a global $U(3)^5$
symmetry,
\begin{equation}
G_{3}\equiv U(3)_q\times U(3)_l\times U(3)_u \times U(3)_d \times U(3)_e\, .
\end{equation}
The introduction of Yukawa interactions in the SM Lagrangian preserves a global hypercharge symmetry $U(1)_Y$, a baryonic symmetry $U(1)_B$, and three leptonic symmetries, $U(1)_e$ $U(1)_\mu$ and $U(1)_\tau$.

Our first approximation when imposing flavor symmetries on SMEFT predictions is to assume that the dimension-6 SMEFT coefficients have the $G_3$ symmetry of the (Yukawa-less) SM. This symmetry prevents the generation of the operators ${\Op}_{uW}[33]$ and ${\Op}_{uB}[33]$ since they carry a left-right fermionic current. 

Under $G_3$, the other operators in Class A respect the form
\bea
{\text{Class~A}}:\quad{\C}_X[ii]=E_X,\,\quad i=1,2,3,
\label{eq:g3a}
\eea
All the 2-fermion operators that contribute to the EWPOs automatically have the structure of Eq. \ref{eq:g3a}, 
and there are $7$ real coefficients in this class, consistent with the counting of Ref. \cite{Faroughy:2020ina}.

Regarding the operators of Class B, both contractions of fermion indices  
\begin{equation}
({\overline{f}}_i\gamma^\mu f_i)({\overline{f}}_j\gamma^\mu f_j),\,\qquad 
({\overline{f}}_i\gamma^\mu f_j)({\overline{f}}_j\gamma^\mu f_i),\,
\end{equation}
are invariant under $U(3)^5$, so the coefficients of Class B reduce to 
\bea
{\text{Class~B}}:\quad{\C}_Y[iiii]&=&A_Y+B_Y,\,{\C}_Y[iijj]=A_Y,\,{\C}_Y[ijji]=B_Y,\,\quad  i\neq j\, \&\, i,j=1,2,3.
\label{eq:g3b}
\eea
Remembering that $B_{ee}=A_{ee}$, there are 11 independent coefficients in Class B. 

Finally, the operators in Class C have the structure
\begin{equation}
{\text{Class~C}}:\quad  {\C}_Z[iijj]=D_Z\quad  i,j=1,2,3,
  \end{equation}
for a total of 11 independent coefficients also in Class C. In total, if we impose a $G_3$ symmetry
on the SMEFT Lagrangian, we are left with 29 independent coefficients contributing to the EWPOs at one loop.\footnote{We note that the (RR)(RR) operator, ${\Op}_{ud}^{(8)}$, does not contribute to EWPOs at one-loop, and so our counting of the $U(1)^5$ operators agrees with Ref. \cite{Greljo:2022cah}.}

  \subsection{Flavor Assumptions: MFV}
The Yukawa couplings break the $U(3)^5$ global symmetry in the SM.  If we assume that this is the ${\it{only}}$
source of breaking of the $G_3$ symmetry, we are led to the MFV scenario\footnote{Since we assume that the CKM matrix is diagonal, there is no flavor violation in our implementation of the MFV scenario.}\cite{DAmbrosio:2002vsn}.  The SM Yukawa couplings are considered as $U(3)^5$ auxiliary fields with the transformations,
\begin{eqnarray}
Y_u&\sim& (3,1,{\overline{3}},1,1)\nonumber \\
Y_d&\sim& (3,1,1,{\overline{3}},1)\nonumber \\
Y_e&\sim &(1,3,1,1,{\overline{3}})\, .
\end{eqnarray}
It is always possible to choose a basis such that
$Y_u,~Y_d$, and $Y_e$ are diagonal matrices.
We remind the reader that we have assumed that the CKM matrix is diagonal and 
the only  non-zero fermion mass is assumed to be the top quark mass, $M_t$.  The coefficients of the operators containing at least 2 top quarks are modified,
while the other coefficients retain the $G_3$ structure. 
 Our implementation of the MFV scenario with a massless $b$ quark is equivalent to a global 
$U(2)_q\times U(3)_l\times U(2)_u \times U(3)_d \times U(3)_e$ symmetry.

In Class A, the only operators with a flavor structure that depends on the third generation fermions are ${\Op}_{\phi u},~{\Op}_{\phi q}^{(3)}$, ${\Op}_{\phi q}^{(1)}$, ${\Op}_{u W}$ and ${\Op}_{u B}$.
\begin{eqnarray}
\begin{array}{l l l l}
{\text{Class ~A}}:  &\quad\C_X[\alpha\alpha]=E_X,\,{\C}_X[33]= E^{(3)}_X,\,
&\quad {\Op}_X\equiv{\Op}_{\phi u}\,,{\Op}_{\phi q}^{(3)},\,{\Op}_{\phi q}^{(1)},\,&\quad  \alpha=1,2
\nonumber \\
&\quad{\C}_{\tilde{X}}[ii]=E_{\tilde{X}}, 
&\quad{\Op}_{\tilde{X}}  \equiv
{\Op}_{\phi e},\,  {\Op}_{\phi d},\, 
 {\Op}_{\phi l}^{(3)} ,\, {\Op}_{\phi l}^{(1)},\,&\quad  i=1,2,3\, .
 \end{array}
\end{eqnarray}
Since for ${\Op}_{u W}$ and ${\Op}_{u B}$ only the $\C_{uW}[33]$ and $\C_{uB}[33]$ enter in our calculations, a total of 12 independent coefficients contribute to the EWPO at NLO, in agreement with Table 9 of \cite{Faroughy:2020ina}.

The operators in Class B involving charge ${2\over 3}$- quarks all have 4 $u_R$ fields or 4 $q_L$ fields.  Retaining the contributions up to 
${\cal{O}}({M_t^2\over v^2})$
,
the coefficients of ${\Op}_{uu}$, ${\Op}_{qq}^{(3)}$ and ${\Op}_{ qq}^{(1)}$ are subject to the following relations in 
the MFV scenario,
\begin{eqnarray}
{\text{Class ~B}}:\quad &&{\C}_Y[1122]={\C}_Y[2211]=A_Y,\,{\C}_Y[1221]={\C}_Y[2112]=B_Y\nn\\
&&{\C}_Y[33\alpha\alpha]={\C}_Y[\alpha\alpha33]=A_Y^{(3)},\,{\C}_Y[3\alpha\alpha3]={\C}_Y[\alpha33\alpha]=B_Y^{(3)}\nn\\
&&{\C}_Y[\alpha\alpha\alpha\alpha]=A_Y+B_Y,\,{\C}_Y[3333]=2(A_Y^{(3)}+B_Y^{(3)})-A_Y-B_Y,\,\quad\alpha=1,2
 \end{eqnarray}
which reduces the number of independent coefficients to 4 for each operator. The remaining operators in Class B satisfy the $U(3)^5$ relations of Eq. \ref{eq:g3b}, thus resulting in 17 independent coefficients in Class B.
 
In Class C, the operators ${\Op}_{lu},~{\Op}_{qe},~{\Op}_{qd}^{(1)},~{\Op}_{eu},~{\Op}_{ud}^{(1)},~{\Op}_{lq}^{(1)}~{\Op}_{lq}^{(3)}$ have 2 top quarks and they satisfy the relations: 
\begin{eqnarray}
{\text{Class~C}}:\quad{\C}_Z[\alpha\alpha ii]=D_Z,\, {\C}_Z[33ii]=D_Z^{(3)},\,\quad\alpha=1,2,\,i=1,2,3
 \end{eqnarray} 
 and hence we have 2 independent coefficients for each operator.

 The only operator in Class C with 2 charge- ${2\over 3}$ quarks contributing to EWPOs is ${\Op}_{qu}^{(1)}$ which satisfies the relations,
 \begin{eqnarray}
{\text{Class~C}}:\quad &&{\C}_{qu}^{(1)}[\alpha\alpha\beta\beta]=D_{qu^{(1)}},\,{\C}_{qu}^{(1)}[3333]=D_{qu^{(1)}}^{(33)}\nonumber \\ 
&&{\C}_{qu}^{(1)}[33\alpha\alpha]=D_{qu^{(1)}}^{(3)},\,{\C}_{qu}^{(1)}[\alpha\alpha33]=D_{qu^{(1)}}^{(\bar{3})},\,\quad\alpha,\beta=1,2,
 \end{eqnarray}
 and hence there are 4 independent flavor structures for $C_{qu}^{(1)}$. The remaining operators in Class C,
 ${\Op}_{ed}, {\Op}_{ld}$ and ${\Op}_{le}$, satisfy the $U(3)^5$ relations and
 contribute 1 independent operator each.
 Overall, we have 21 independent coefficients in Class C.
Excluding the operators ${\Op}_{qd}^{(8)},~{\Op}_{qu}^{(8)},$ and ${\Op}_{ud}^{(8)}$ which
do not contribute to EWPOs at NLO, our counting is in agreement with Table 2  of  \cite{Bruggisser:2021duo} and Table 9 of \cite{Faroughy:2020ina}.

\subsection{Flavor Assumptions: $U(2)^5$}
Here we consider a scenario where the new physics distinguishes between the $3^{rd}$ generation
and the first 2 generations. 
The first 2 generations are assumed to have the global symmetry\cite{Greljo:2022cah},
\begin{equation}
G_2\equiv U(2)_q\times U(2)_l\times U(2)_u \times U(2)_d \times U(2)_e\, .
\end{equation}
Since we consider only the top quark mass to be non-zero, the $U(2)^5$ symmetry is unbroken.

Using the classification of operators given in Sec. \ref{sec:g3},
the $G_2$ symmetry requires that operators of Class A have the form:
\begin{eqnarray}
{\text{Class~A}}:\quad &&{\C}_X[\alpha\alpha]=E_X,\nonumber \\
&&{\C}_X[33]=E_X^{(3)},\,\quad\alpha=1,2.
\label{eq:g2a}
\end{eqnarray}
 There are $16$ independent coefficients in this class.

The Class B operators satisfy
\begin{eqnarray}
\begin{array}{l l l}
{\text{Class~B}}:&\quad{\C}_Y[\alpha\alpha\beta\beta]=A_Y,\,{\C}_Y[\alpha\beta\beta\alpha]=B_Y,\,{\C}_Y[\alpha\alpha\alpha\alpha]=A_Y+B_Y,&\\
&\quad{\C}_Y[\alpha\alpha33]={\C}_Y[33\alpha\alpha]=A_Y^{(3)},\,{\C}_Y[3\alpha\alpha3]={\C}_Y[\alpha33\alpha]=B_Y^{(3)},&\\
&\quad{\C}_Y[3333]=F_Y^{(3)},&\quad\alpha\neq\beta,\,\alpha,\beta=1,2,
\end{array}
  \end{eqnarray}
so that each operator has 5 independent coefficients. As before, ${\Op}_{ee}$ has an extra constraint coming from the Fierz identities, which reduces the number of independent coefficients to 3 for this operator. Therefore, since ${\Op}_{uu}[3333]$ does not contribute, Class B has 27 coefficients contributing to EWPOs at NLO in total. 
  
The Class C operators, $\Op_C$, contain 2 different fermion representations and the $U(2)^5$ symmetry implies,
 \begin{eqnarray}
{\text{C}}:\quad &&{\C}_Z[\alpha\alpha\beta\beta]=D_Z,\,{\C}_Z[3333]=D_Z^{(33)}\nonumber \\ 
&&{\C}_Z[33\alpha\alpha]=D_Z^{(3)},\,{\C}_Z[\alpha\alpha33]=D_Z^{(\bar{3})},\,\quad\alpha,\beta=1,2,
 \end{eqnarray}
 and there are 4 independent coefficients corresponding to each operator structure, for a total
 of 44 operators.

\subsection{Third Generation Centric}
In this scenario, only operators involving  the $3^{rd}$ generation fermions are non-zero.  An example of
such a scenario might be a $Z^\prime$ boson that only interacts with the $3^{rd}$ generation.
Operators of Class A in this scenario have the form:
\begin{eqnarray}
{\text{Class~A}}:\quad &&{\C}_X[\alpha\alpha]=0, \nonumber \\
&&{\C}_X[33]=E_X^{(3)},\,\quad \alpha=1,2.
\label{eq:tca}
\end{eqnarray}
 There are $9$ real coefficients in this class.
 
 The Class B and C operators satisfy
\begin{eqnarray}
\begin{array}{l l l}
{\text{Class~B~\&~C}}:&\quad{\C}_{Y,Z}[\alpha\alpha\beta\beta]={\C}_Y[\alpha\beta\beta\alpha]=0,& \\
&\quad{\C}_{Y,Z}[\alpha\alpha33]={\C}_{Y,Z}[33\alpha\alpha]={\C}_Y[3\alpha\alpha3]={\C}_Y[\alpha33\alpha]=0,&\\
&\quad{\C}_Y[3333]=F_Y^{(3)},\,{\C}_Z[3333]=D_Z^{(3)},\,&\quad \alpha,\beta=1,2.
\end{array}
  \end{eqnarray}
For a total of 5 coefficients in Class B and 11 coefficients in Class C.

\subsection{Third Generation Phobic}

In this scenario, we assume that the new physics only couples to the $1^{st}$ 2 generations.  An example of such a model can be found in \cite{Crivellin:2021bkd}.
We assume no further symmetry, but note that the operators contributing to EWPOs do not have an arbitrary flavor structure, as described at the beginning of this section.  

The flavor structure of the operators of Class A is
\begin{eqnarray}
{\text{Class~A}}:\quad &&{\C}_X[\alpha\alpha]=E_X^{(\alpha)}. \nonumber \\
&&{\C}_X[33]=0,\,\qquad \alpha=1,2,
\label{eq:tpa}
\end{eqnarray}
for a total of 14 independent coefficients.

 For Class B we have
\begin{eqnarray}
{\text{Class~B}}:\quad&&{\C}_Y[\alpha\alpha\alpha\alpha]=F_Y^{(\alpha)},\,{\C}_Y[\alpha\alpha\beta\beta]=A_Y^{(\alpha\beta)},\,{\C}_Y[\alpha\beta\beta\alpha]=B_Y^{(\alpha\beta)},
\nonumber \\
&&{\C}_Y[\alpha\alpha33]={\C}_Y[33\alpha\alpha]={\C}_Y[3\alpha\alpha3]={\C}_Y[\alpha33\alpha]={\C}_Y[3333]=0,\,\quad \alpha\neq\beta,\,\alpha,\beta=1,2,
  \end{eqnarray}
corresponding to 4 independent coefficients for each operator (3 for ${\Op}_{ee}$), for a total of 23 coefficients in Class B.

Finally for Class C the available structures are
\begin{eqnarray}
{\text{Class~C}}: \quad&&{\C}_Z[\alpha\alpha\beta\beta]=D_Z^{(\alpha\beta)}\nonumber \\
&&{\C}_Z[\alpha\alpha33]={\C}_Z[33\alpha\alpha]={\C}_Z[3333]=0,\,\quad \alpha,\beta=1,2,
  \end{eqnarray}
which again gives 4 independent coefficients for each operator, for a total of 44 coefficients in Class C.

\subsection{Third Generation Phobic + $U(2)^5$}

As a specific case of the previous example, we consider a scenario where, after assuming that new physics couples only to the $1^{st}$ 2 generations, we further assume the existence of a $U(2)^5$ symmetry. 

In this case, the flavor structure of the operators of Class A is
\begin{eqnarray}
{\text{Class~A}}:\quad &&{\C}_X[\alpha\alpha]=E_X, \nonumber \\
&&{\C}_X[33]=0,\,\qquad \alpha=1,2,
\label{eq:tpa2}
\end{eqnarray}
for a total of 7 independent coefficients.

 For Class B we have
\begin{eqnarray}
{\text{Class~B}}: \quad&&{\C}_Y[\alpha\alpha\beta\beta]=A_Y,\,\quad{\C}_Y[\alpha\beta\beta\alpha]=B_Y,\,{\C}_Y[\alpha\alpha\alpha\alpha]=A_Y+B_Y,
\nonumber \\
&&{\C}_Y[\alpha\alpha33]={\C}_Y[33\alpha\alpha]={\C}_Y[3\alpha\alpha3]={\C}_Y[\alpha33\alpha]={\C}_Y[3333]=0,\,\quad \alpha\neq\beta,\,\alpha,\beta=1,2,
  \end{eqnarray}
corresponding to 2 independent coefficients for each operator (1 for ${\Op}_{ee}$), for a total of 11 coefficients in Class B.

Finally for Class C the available structures are
\begin{eqnarray}
{\text{Class~C}}: \quad&&{\C}_Z[\alpha\alpha\beta\beta]=D_Z\nonumber \\
&&{\C}_Z[\alpha\alpha33]={\C}_Z[33\alpha\alpha]={\C}_Z[3333]=0,\,\quad\alpha,\beta=1,2,
  \end{eqnarray}
which again gives 1 independent coefficient for each operator, for a total of 11 coefficients in Class C.

\subsection{Flavorless Scenario}
This is the scenario employed in our previous calculations\cite{Dawson:2019clf}, where flavor plays no role and we simply
drop all flavor indices.   In this scenario, we assume that the operators have no flavor structure, that is we make the replacements,
\bea
{\C}_{X,Y,Z}[\dots]=A_{X,Y,Z}\, .
\eea
Therefore in this case there are 26 coefficients that contribute. 

We summarize the results of our discussion of flavor scenarios in Table \ref{tab:opflav}.  The number of operators contributing to EWPOs at NLO varies dramatically depending on the flavor structure and is significantly smaller than the 178 independent coefficients that contribute with no
assumptions about flavor.  We will see that these choices have large effects on the numerical results of fits to EWPOs.
\begin{table}[t] 
\centering
\renewcommand{\arraystretch}{1.5}
\begin{tabular}{||c||c|c|c|c|c|c|c||} 
\hline \hline
Operator & $U(3)^5$ & MFV &$U(2)^5$ &$3^{rd}$ gen specific & $ 3^{rd}$ gen phobic & $ 3^{rd}$ gen phobic + $U(2)^5$& Flavorless\\ 
\hline\hline
Class A &7&12&16&9&14&7&9
\\
\hline
Class B &11&17&27&5&23&11&6
\\
\hline
Class C&11&21&44&11&44&11&11
\\
\hline \hline
Total&29&50&87&25&81&29&26\\
\hline \hline
\end{tabular}
\caption{Number of independent operators contributing to NLO predictions for EWPOs in various flavor scenarios.\label{tab:opflav}}
\end{table}

\section{NLO}
\label{sec:nlo}

This section describes our calculational procedure for the NLO corrections to the electroweak precision observables. Some of the results can be obtained from our previous calculation\cite{Dawson:2019clf} by generalizing the flavor structure of each observable. However, including a general flavor structure at NLO requires a new calculation for many of the flavor structures
and observables. Therefore, we have performed the calculation  in all generality and in this
section we describe the details of the calculation and the assumptions that are 
implicit in the numerical results of the next section.

\begin{figure}[t]
	\centering
\includegraphics[width=2in]{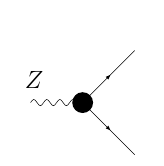}  
	\caption{Z decays to fermions.  The black circle represents the tree level SMEFT interaction. }
	\label{fig:zdec}
\end{figure}
At tree level, the interactions are just the $Z$ (and $W$)  decays to 2 fermions shown in Fig. \ref{fig:zdec}
and the effect of including the dimension-6  SMEFT contributions
is to change the vector boson couplings to fermions.
The tree level SMEFT couplings of fermions  to the $Z$ and $W$ are given  in terms of our input parameters ($\alpha, M_Z, G_\mu$),
\begin{eqnarray}
  L&\equiv&  2M_Z \sqrt{\sqrt{2}G_\mu }\biggl\{Z_\mu\biggl[g_L^{Zq}+\delta g_{L}^{Zq}\biggr]
  {\overline q}\gamma_\mu q
 +  Z_\mu\biggl[g_R^{Zu}+\delta g_{R}^{Zu}\biggr]{\overline u}_R\gamma_\mu u_R
 \nonumber \\ &&\mathrel{\hphantom{2M_Z \sqrt{\sqrt{2}G_\mu }}}+
 Z_\mu\biggl[g_R^{Zd}+\delta g_{R}^{Zd}\biggr]{\overline d}_R\gamma_\mu d_R +
Z_\mu\biggl[g_L^{Zl}+\delta g_{L}^{Zl}\biggr]
  {\overline l}\gamma_\mu l \nonumber \\ &&\mathrel{\phantom{2M_Z \sqrt{\sqrt{2}G_\mu }}} 
 +  Z_\mu\biggl[g_R^{Ze}+\delta g_{R}^{Ze}\biggr]
  {\overline e}_R\gamma_\mu e_R
  +  Z_\mu\biggl( \delta g_{R}^{Z\nu}\biggr)
  {\overline \nu}_R\gamma_\mu \nu_R\biggr\}
  \nonumber \\
&  +&
 2M_W[\alpha, M_Z, G_\mu] \sqrt{\frac{G_\mu}{\sqrt{2}} }
\biggl\{W_\mu\biggl[(1+\delta g_{L}^{Wq}){\overline u}_L\gamma_\mu d_L
  +\biggl(\delta g_R^{Wq}\biggr) 
  {\overline u}_R\gamma_\mu d_R\biggr] 
  \nonumber \\
  &&\mathrel{\phantom{2M_Z \sqrt{\sqrt{2}G_\mu }}} +W_\mu\biggl[(1+\delta g_{L}^{Wl}){\overline \nu}_L\gamma_\mu e_L
  \biggr] 
  +h.c.\biggr\}\, ,
  \label{eq:gldef}
  \end{eqnarray}
  where at tree level,
\begin{eqnarray}
g_R^{Zf}&=&-s_W^2 Q_f\quad{\rm and}\quad g_L^{Zf}=T_3^f -s_W^2 Q_f\nonumber \\
 s_W^2&\equiv & 1-\frac{M_W^2[\alpha, M_Z, G_\mu]}{M_Z^2}\, ,
\end{eqnarray}
$T_3^f=\pm \displaystyle \frac{1}{2}$, and $M_W^2[\alpha, M_Z, G_\mu]$ is the mass of the $W$ boson written in terms of the ($\alpha, M_Z, G_\mu$) input parameters as defined later in Sec. \ref{subs:MW}.
Analytic expressions for the shifts in the couplings in terms of the Warsaw basis coefficients can be found in \cite{Baglio:2017bfe,Brivio:2017bnu}.

We use the SMEFT Feynman rules of Ref.  \cite{Dedes:2017zog} as implemented in FeynRules \cite{Alloul:2013bka} from which a FeynArts  \cite{Hahn:2000kx}  model file is obtained. In this way, we are able to generate  the relevant one-loop and real emission amplitudes, which include all the QCD and EW corrections to the $Z$ and $W$ pole observables considered here. 
All the amplitudes are generated in $R_\xi$ gauge, and the gauge invariance of the final results is verified explicitly. The results of the previous step are manipulated with FeynCalc \cite{Mertig:1990an,Shtabovenko:2016sxi}, which is also used to reduce the integrals that appear in terms of Passarino-Veltman integrals \cite{Passarino:1978jh}. We retain only terms of ${\cal{O}}(v^2/\Lambda^2)$, since the SMEFT is renormalizable at fixed order in the $(v^2/\Lambda^2)$ expansion\footnote{Including terms resulting from squaring the LO EFT result would result in contributions of ${\cal{O}}(v^4/\Lambda^4)$, which is the same order as the dimension-8 terms we are neglecting, and hence would be inconsistent.} . As is the case of the pure SM results, the Feynman diagrams that contribute to the calculations are afflicted by both UV and IR divergences and we describe the regulation of these singularities below.

\subsection{Real contributions}

\begin{figure}[t]
	\centering
\includegraphics[width=1.5in]{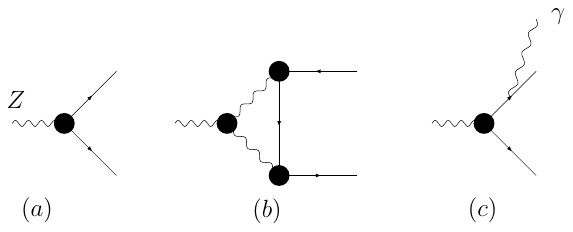}  
	\caption{Z decays to 2 fermions, including real photon emission.  The black circle represents the tree level SMEFT interaction. }
	\label{fig:zreal}
\end{figure}

The decays of the $Z$ and $W$ bosons receive contributions from real photon and gluon emissions (as demonstrated generically in Fig. \ref{fig:zreal}), which are necessary to cancel the IR divergences from the virtual contributions. From Fig. \ref{fig:zreal}, it is clear that the flavor structure of the real emission diagrams is identical to that of the tree level result. Due to the  complicated tensor structures of the SMEFT vertices, it is  convenient to avoid integration over the final state 3-body phase space and  therefore we use the method of Ref. \cite{Anastasiou:2002yz},  replacing the integration over the phase space with a loop integration. Specifically, we square the tree-level real amplitudes and then make use of the identity, 
\bea
2i\pi\delta(p^2-m^2)=\frac1{p^2-m^2+i0}-\frac1{p^2-m^2-i0}\, ,
\label{eq:Cutkosky}
\eea
to build a correspondence between phase space integrals and the imaginary part of 2-loop integrals, which we calculate by rewriting them in terms of known master integrals \cite{Tarasov:1997kx, Martin:2003qz} using FIRE \cite{Smirnov:2014hma}. We notice that only the imaginary parts that correspond to a cut through all three particles in the final state contribute to the phase space integral, and thus particular attention has to be taken in isolating the correct imaginary part.
We then check that the results obtained with this technique reproduce the known results for the real photon and gluon contributions to the SM NLO results. Finally, we treat IR divergences in the real and virtual contributions by giving a small mass to photons and gluons, and then carefully taking the massless limit of the full result\cite{Hollik:1988ii}. 

\subsection{Virtual contributions}

\begin{figure}[t]
	\centering
\includegraphics[width=1.5in]{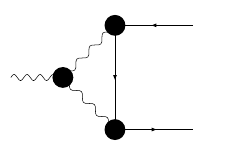}
\includegraphics[width=1.5in]{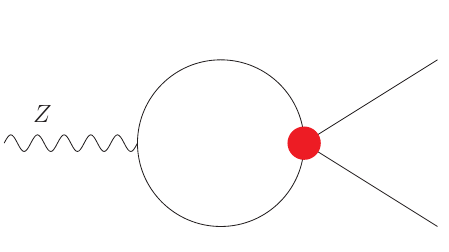}    
	\caption{Sample virtual one-loop contributions to Z decays to 2 fermions.  The black circles represent SMEFT insertions correcting the SM vertices, while the red circle denotes the dimension-6 4-fermion operators. }
	\label{fig:zvirt}
\end{figure}

At LO, the amplitudes can be written  in terms of the EW vacuum expectation value ($v$), the pole masses of the $Z$ and $W$ bosons, and the Wilson coefficients of SMEFT operators in Table \ref{tab:opdef}.  Tree level results for the observables of Eq. \ref{eq:quan} are given in Appendix A of \cite{Dawson:2019clf}.  This reference also includes the NLO results in the case where
the 2-fermion and 4-fermion operators are assumed to be independent of flavor.   The virtual diagrams consist of the full set of one-loop EW and QCD diagrams with the SMEFT vertices. Contributions of special interest to the study of the flavor structure arise from the 4-fermion interactions shown in Fig. \ref{fig:zvirt}.  
In what follows, we do not assume any flavor structure for the 4 fermion operators, although as noted in Sec. 
\ref{sec:smeft}, only certain flavor structures arise. 

\subsection{UV counterterms}

To treat the UV divergences we use dimensional regularization in $d$ dimensions so that the poles are written in terms of the regulator $\epsilon=(4-d)/2$, which are later canceled by adding appropriate counterterms. 

Note that we chose to exactly cancel all the tadpole contributions, that is we put to zero from the beginning  all the diagrams that contain tadpoles. This corresponds to identifying the renormalized vacuum with the (gauge-dependent) minimum of the renormalized Higgs potential\cite{Degrassi:2014sxa}. Consequently, all intermediate quantities are in general gauge dependent and only the final physical observables are gauge independent. Therefore, the cancellation of the $R_\xi$ gauge parameter is highly non trivial and a strong check of the correctness of the calculation.

Since we do not have direct measurements of the coefficients of the dimension-6 operators, we treat them as $\overline{{MS}}$ parameters by constructing renormalized coefficients, $\C_i(\mu)$, of the form \beq
\C_i(\mu)=\C_{0,i}+\delta\C_i=\C_{0,i}-\frac1{2\hat{\epsilon}}\frac1{16\pi^2}\gamma_{ij}\C_j,
\eeq
where we define $\hat{\epsilon}^{-1}\equiv\epsilon^{-1}-\gamma_E+\log(4\pi)$ . In the previous equation, $\mu$ is the renormalization scale, $\C_{0,i}$ are the bare quantities, and  $\gamma_{ij}$ are the one-loop anomalous dimensions defined by the relation
\beq
\mu \frac{d \C_i}{d\mu}=\frac1{16\pi^2}\gamma_{ij}\C_j.
\eeq
The complete set of anomalous dimensions for dimension-6 SMEFT operators in the Warsaw basis is in Refs. \cite{Jenkins:2013zja,Jenkins:2013wua,Alonso:2013hga}. 
 In
general the $\gamma_{ij}$ are not diagonal and 4 -fermion operators mix with 2-fermion 
and other operators under renormalization.

Regarding the SM input parameters, we use the on-shell (OS) scheme which defines   the renormalized quantities in
terms of measured observables, and specifically we choose  $\{\alpha,G_\mu,M_Z\}$ as the independent inputs\cite{Brivio:2017bnu,Brivio:2021yjb}. 

The vacuum expectation value $v$ is connected to the Fermi constant $G_\mu$ by the relation,
\beq
G_\mu+{1\over\sqrt{2}\Lambda^2}\C_{ll}[1221]-{1\over\sqrt{2}\Lambda^2}(\C_{\phi l}^{(3)}[11]+\C_{\phi l}^{(3)}[22])\equiv\frac1{\sqrt{2} v_{0}^2}(1+\Delta r),
\label{eq:geftdef}
\eeq
where 
the $\C$ in Eq. \ref{eq:geftdef} are the renormalized $\C_i(\mu)$,
$v_0$ is the unrenormalized minimum of the potential (where we use the notation '0' to denote the tree level
relation), and $\Delta r$ is extracted from the calculation of  muon decay at 1-loop at zero momentum transfer. The
result for $\Delta r$ in terms of the dimension-6 coefficients is in Appendix D of \cite{Dawson:2018pyl}.

For the mass of the $Z$ boson, the relation between the bare and renormalized quantity is obtained from the formula
\beq
M_Z^2=M^2_{0,Z}-\Pi_{ZZ}(M^2_{Z}),
\eeq
where $M^2_{0,Z}$ is the bare quantity and $\Pi_{ZZ}(M^2_Z)$ is the one-loop correction to the 2-point function for Z, which is reported analytically for the SMEFT in Refs.  \cite{Chen:2013kfa,Ghezzi:2015vva}. It is also convenient to define the one-loop corrections to the W boson 2-point function $\Pi_{WW}(M^2_W)$ in a similar manner.

The relation between the bare and renormalized electromagnetic coupling $\alpha$ is 
\beq
\alpha(M_Z)=\frac{\alpha}{1-\Delta\alpha(M_Z)} \, ,
\eeq
where we derive $\Delta\alpha$ from the 1-loop SMEFT corrections to the $\gamma \bar{e}e$ vertex. Note that $\Delta\alpha$ receives non-perturbative corrections from diagrams involving light quarks that contribute to the vacuum polarization $\Pi_{\gamma\gamma}^{(5)}(0)$. To treat these contributions we use the hadronic contribution to $\Delta\alpha$ which is defined as
\beq
\Delta\alpha_{\rm had}^{(5)}(M_Z)=4\pi\alpha\left(\Pi^{(5)}_{\gamma\gamma}(M_Z)-\Pi^{(5)}_{\gamma\gamma}(0)\right),
\eeq
where $\Pi_{\gamma\gamma}^{(5)}(M_Z)$ is a perturbative correction obtained by calculating the same diagrams that contribute to $\Pi_{\gamma\gamma}^{(5)}(0)$. The value of the hadronic contribution to $\Delta\alpha$ is obtained using dispersion relations and experimental data and we report it in Eq. \ref{eq:inputs} together with other inputs.

Finally, it is convenient to define the quantities
\bea
&&\Delta\rho\equiv\Pi_{WW}(M^2_W)/M_W^2-\Pi_{ZZ}(M^2_Z)/M_Z^2\,\,\,\,\,{\rm and}\\
&&\Delta r_{W}\equiv\Delta r+\Pi_{WW}(M^2_W)/M_W^2,
\eea
which will enter in some of the definitions in the following section. 

\subsection{$M_W$ calculation}
\label{subs:MW}

The relation between the mass of the $W$ boson and the input parameters in the SM is given by the well-known formula
\beq
M_{W,SM}^2=M_Z^2\left(1+\sqrt{1-\frac{4\pi\alpha(1+\Delta \bar{r}_{SM})}{\sqrt{2}M_Z^2 G_\mu}}\right),
\label{eq:MWdefSM}
\eeq
where we use a slightly unusual notation to distinguish between the finite gauge-invariant $\Delta \bar{r}$ and  the UV divergent gauge-dependent $\Delta r$. $\Delta \bar{r}_{SM}$ is obtained from the previous corrections:
\beq
\Delta \bar{r}_{SM}=\Delta\alpha_{SM}-\frac{\cos^2\theta_W}{\sin^2\theta_W}\Delta\rho_{SM}+\Delta r_{W,SM},
\eeq

where $\theta_W$ is the weak mixing angle, which is defined in the OS scheme as $\cos\theta_W\equiv M_W/M_Z$. 

In the SMEFT, Eq.  \ref{eq:MWdefSM} is modified:
\beq
M_{W}^2=M_Z^2\left(1+\sqrt{1-\frac{4\pi\alpha(1+\Delta \bar{r})}{\sqrt{2}M_Z^2 G_\mu}}-\frac{1}{16} \frac{1}{\sqrt{2}G_\mu\Lambda^2}\left(\frac{\cos\theta_W}{\cos2\theta_W}\right)^3(F_1G_1+F_2 G_2)\right) ,
\label{eq:MWdef}
\eeq
where 
\bea
\Delta \bar{r}&=&\Delta\alpha-\frac{\cos^2\theta_W}{\sin^2\theta_W}\Delta\rho+\Delta r_{W}\\
&+&\frac{1}{\sqrt{2}G_\mu\Lambda^2}\left(-\frac{\cos\theta_W(\cos\theta_W C_{\phi D}+2\sin\theta_W C_{\phi WB})}{2\sin^4\theta_W}\Delta\rho_{SM}\right.\nn\\
&+&\frac{\cos\theta_W (\cos\theta_W C_{\phi D}+4\sin\theta_W C_{\phi WB})}{2\sin^2\theta_W}\left(\Delta r_{W,SM}-\frac{\Pi_{WW,SM}(M^2_W)}{M_W^2}\right)\nn\\
&+&\left.\frac{\cos^2\theta_W}{2\sin^2\theta_W}\delta\C_{\phi D}+\frac{2\cos\theta_W}{\sin\theta_W}\delta\C_{\phi WB}\right),\nn
\eea
and we define
\bea
G_1&=& \cos\theta_W \C_{\phi D}+4\sin\theta_W \C_{\phi WB}+\sin\theta_W\tan\theta_W
(2\C_{\phi l}^{(3)}[11]+2\C_{\phi l}^{(3)}[22]-2C_{ll}[1221])\\
G_2&=&\cos\theta_W \C_{\phi D}+2\sin\theta_W \C_{\phi WB}\\
F_1 &=& 4 + 6 \Delta r_{SM} + 2 (2 + \Delta r_{SM}) \cos4\theta_W - M_Z^2 \frac{\partial\Delta r_{SM}}{\partial (M_W^2)} \sin2\theta_W \sin4\theta_W\\
F_2 &=&-8\Delta r_{SM}\cos2\theta_W.
\eea

It is convenient to rewrite Eq. \ref{eq:MWdef} as,
\beq
M_W^2=M_{W,SM}^{2}+ \frac{1}{\sqrt{2}G_\mu\Lambda^2} M_{W,EFT}^{2}
\eeq
where $M_{W,SM}^{2}$ is obtained from Eq. \ref{eq:MWdefSM}, and we separate the SM and EFT contributions. At  NLO, the mass of the $W$ boson is
\beq
M_{W,NLO}=\sqrt{M_{W,SM}^{2}+ \frac{1}{\sqrt{2}G_\mu\Lambda^2} M_{W,EFT}^{2}}\eqsim\sqrt{M_{W,SM}^{2}}+ \frac{1}{\sqrt{2}G_\mu\Lambda^2} \frac{M_{W,EFT}^{2}}{2\sqrt{M_{W,SM}^{2}}}\, ,
\label{eq:MWNLO}
\eeq
and we calculate $M_{W,SM}^{2}$ and $M_{W,EFT}^{2}$ at 1-loop. Note that the calculation of $M_{W,SM}^{2}$ and $M_{W,EFT}^{2}$ involve the mass of the $W$ itself that enters through $\cos\theta_W$. In this case, we use the value of $M_W$ obtained from Eq. \ref{eq:MWdefSM} after putting $\Delta \bar{r}\to0$. As the last step in our calculation, we replace the SM result
from our calculation  with the "best" theory result of Table \ref{tab:expnums}, obtaining our final expression for the mass of the $W$ boson:
\beq
M_{W,fin}= M_{W,"best"}+ \frac{1}{\sqrt{2}G_\mu\Lambda^2} \frac{M_{W,EFT}^{2}}{2\sqrt{M_{W,SM}^{2}}}.
\label{eq:MWbest}
\eeq

All the other observables that we include in our fit depend on the mass of the $W$ boson in some way. In our numerical calculation of a generic observable, $O$, we use our best knowledge of $M_W$ (Eq. \ref{eq:MWbest}) in the calculation of the LO contributions to $O$ and $M_{W,NLO}$ in Eq. \ref{eq:MWNLO} in the calculation of the NLO contributions to $O$. Afterwards, we replace our SM result with the "best" theory results in table \ref{tab:expnums}. Finally, our EWPOs are expressed as

\beq
O=O_{"best"}+O_{EFT,LO}|_{M_W\to M_{W,fin}}+O_{EFT,NLO}|_{M_W\to M_{W,NLO}}\, .
\eeq

We close this section with a few remarks about the validity  of our expansion.  After renormalization, the amplitude is finite.  We then square the amplitude and retain only terms that are linear in the SMEFT coefficients.  In principle, we could retain the terms quadratic in the SMEFT coefficients as this is also  finite.  However, the quadratic contributions generically have the same counting in factors of ${1\over 16\pi^2}$ as does a 2-loop SMEFT contribution with 2 insertions of SMEFT coefficients interfering with the SM amplitude.  Furthermore, the quadratic terms contain contributions from SMEFT operators that we have dropped because they don't interfere with the 
SM.\footnote{ The quadratic contributions corresponding to the square of the amplitude computed here can be obtained from the authors on request.}

\section{Results}
\label{sec:results}
In this section, we show the effects of the flavor assumptions on  global fits to EWPOs.  We fit to the data given in Table \ref{tab:expnums} and we use the Particle Data Group review\cite{PhysRevD.98.030001} value for $M_W$.  The column labelled "best" theory corresponds to predictions computed to the highest available order in perturbation theory. We take as our   input parameters,
 \begin{eqnarray}
G_\mu&=&1.1663787(6)\times 10^{-5} \gev^{-2}\nonumber \\
M_Z&=&91.1876\pm .0021\gev\nonumber \\
{1\over \alpha} &=& {137.035999139(31)} \nonumber\\
\Delta\alpha_{\rm had}^{(5)} &=& 0.02768\pm 0.00009\nonumber\\
\alpha_s(M_Z)&=&0.1181\pm 0.0011 \nonumber \\
M_h&=&125.25\pm 0.17 ~\gev\nonumber\\
M_t&=&172.69\pm0.5~\gev\nonumber\, .
\label{eq:inputs}
\end{eqnarray}
    The SMEFT scale is always taken to be $\Lambda=1~$TeV and the coefficient functions are evaluated at the renormalization scale $\mu=M_Z$.

 Flavor blind predictions for EWPOs were presented in Ref. \cite{Dawson:2022bxd}, which corresponds to the scenario labeled flavorless here.{\footnote{The exact numerical results of this work differ slightly from Ref. \cite{Dawson:2022bxd} since we fit to an expanded set of measurements with slightly different numerical values for the inputs.}}  In this section, we concentrate on the derivation of limits on SMEFT coefficients that are dependent on the assumed flavor scenario.
 Our limits are obtained by considering only one non-zero operator type, $\Op [ijkl]$,  at a time and imposing the flavor symmetries described in Sec. \ref{sec:smeft}.  The single parameter limits
 are  marginalized over the other independent flavor structures. Before proceeding, we find it convenient to briefly describe the marginalization procedure we  use to obtain our results. Assuming Gaussian uncertainties, we calculate the $\chi^2$ function 
 \beq
 \chi^2(\mathcal{C})=\sum_{i,j}(O^{\rm exp}_i-O^{\rm th}_i)\sigma^{-2}_{ij}(O^{\rm exp}_j-O^{\rm th}_j),
 \eeq
 where $O^{\rm exp}_i$ and $O^{\rm th}_i$ are the experimental and theoretical values of a certain observable $O_i$ respectively and  $\sigma^2_{ij}$ is the covariance matrix.
 In general, we can write $\chi^2(\mathcal{C})\equiv\chi^2(\bar{\mathcal{C}},\hat{\mathcal{C}})$ as a function of coefficients $\bar{\mathcal{C}}$, which we are interested in finding the limits of, and coefficients $\hat{\mathcal{C}}$, which we want to marginalize over.  As a first step, we find the values of $\hat{\mathcal{C}}$ that minimize $\chi^2$ while keeping $\bar{\mathcal{C}}$ free. In other words, we find the $\hat{\mathcal{C}}_M(\bar{\mathcal{C}})$ that satisfy the system of equations 
 \beq
 \left.\frac{\partial \chi^2}{\partial\hat{\mathcal{C}}}\right |_{\hat{\mathcal{C}}=\hat{\mathcal{C}}_M(\bar{\mathcal{C}})}=0,
 \eeq
 where we highlighted that the $\hat{\mathcal{C}}_M$ are defined as function of the free parameters $\bar{\mathcal{C}}$.
We can now define, 
\beq
\chi^2(\bar{\mathcal{C}})\equiv\chi^2(\bar{\mathcal{C}},\hat{\mathcal{C}}=\hat{\mathcal{C}}_M).
\eeq
The new function $\chi^2(\bar{\mathcal{C}})$ is used to perform the necessary studies. More details on this procedure can be found in the literature (see for example \cite{Espinosa:2012vu,Berthier:2015gja,Ellis:2018gqa}).
 
For example, we can study the $\chi^2$ for the coefficient $C^{(1)}_{qq}$ in the MFV scenario. The  $\chi^2$ function is given by:
\begin{align}
\chi^2_{\text{MFV}} {\small \left(C_{\text{qq}}^{\text{(1)}} \right)}  & = 21.0184 \ -  0.183188 \ C_{\text{qq}}^{\text{(1)}}[1,1,2,2]  +  0.00528431 \
   C_{\text{qq}}^{\text{(1)}}[1,1,2,2]{}^2  -  0.193834 \
   C_{\text{qq}}^{\text{(1)}}[1,1,3,3] \nonumber  \\ &  +  0.00725802 \
   C_{\text{qq}}^{\text{(1)}}[1,1,2,2]
   C_{\text{qq}}^{\text{(1)}}[1,1,3,3]   +  0.120819 \
   C_{\text{qq}}^{\text{(1)}}[1,1,3,3]{}^2  +  5.33703 \
   C_{\text{qq}}^{\text{(1)}}[1,3,3,1] \nonumber  \\ &  -  0.2973 \
   C_{\text{qq}}^{\text{(1)}}[1,1,2,2]
   C_{\text{qq}}^{\text{(1)}}[1,3,3,1] -  0.271204 \
   C_{\text{qq}}^{\text{(1)}}[1,1,3,3]
   C_{\text{qq}}^{\text{(1)}}[1,3,3,1]  +  4.19859 \
   C_{\text{qq}}^{\text{(1)}}[1,3,3,1]{}^2 \nonumber \\ &  - 5.92823 \
   C_{\text{qq}}^{\text{(1)}}[3,3,3,3]  +  0.163898 \
   C_{\text{qq}}^{\text{(1)}}[1,1,2,2]
   C_{\text{qq}}^{\text{(1)}}[3,3,3,3]  +  0.0186148 \
   C_{\text{qq}}^{\text{(1)}}[1,1,3,3]
   C_{\text{qq}}^{\text{(1)}}[3,3,3,3] \nonumber \\ & -  4.83774 \
   C_{\text{qq}}^{\text{(1)}}[1,3,3,1]
   C_{\text{qq}}^{\text{(1)}}[3,3,3,3]  + 3.85299 \
   C_{\text{qq}}^{\text{(1)}}[3,3,3,3]{}^2.
\end{align}
Following the above procedure, to marginalize $  C_{\text{qq}}^{\text{(1)}}[1,1,2,2]$
  and $C_{\text{qq}}^{\text{(1)}}[3,3,3,3]$, we have
 \beq 
\begin{cases}
\frac{\partial \chi^2_{\text{MFV}}}{\partial C_{\text{qq}}^{\text{(1)}}[1,1,2,2]} & =0 \\   \frac{\partial \chi^2_{\text{MFV}}}{\partial C_{\text{qq}}^{\text{(1)}}[3,3,3,3]}     & =0
\end{cases} \hspace{0.5cm} \Longrightarrow \hspace{0.5cm} 
\begin{array}{l}
   C_{\text{qq}}^{\text{(1)}}[1,1,2,2]
   \to 8.062020\, -0.9688560 \ C_{\text{qq}}^{\text{(1)}}[1,1,3,3]+27.44810 \
   C_{\text{qq}}^{\text{(1)}}[1,3,3,1] \\ 
   \vspace{-0.18cm}
   \\
C_{\text{qq}}^{\text{(1)}}[3,3,3,3]\to 0.597832\, +0.0181909 \
   C_{\text{qq}}^{\text{(1)}}[1,1,3,3]+0.0439991 \
   C_{\text{qq}}^{\text{(1)}}[1,3,3,1]

\end{array}.
\eeq 
Setting the marginalized coefficients to their minimum we find:
\begin{align}
\chi^2_{\text{MFV}} \left( C_{\text{qq}}^{\text{(1)}}[1,1,3,3],C_{\text{qq}}^{\text{(1)}}[1,3,3,1] \right) & =  18.5079\, -0.124191 C_{\text{qq}}^{\text{(1)}}[1,1,3,3]+0.117473
   C_{\text{qq}}^{\text{(1)}}[1,1,3,3]{}^2+0.048042
   C_{\text{qq}}^{\text{(1)}}[1,3,3,1] \nonumber \\  & -0.0711661 \
   C_{\text{qq}}^{\text{(1)}}[1,1,3,3]
   C_{\text{qq}}^{\text{(1)}}[1,3,3,1]+0.012012
   C_{\text{qq}}^{\text{(1)}}[1,3,3,1]{}^2. 
\end{align}

The new $\chi^2$, that now depends only on $C_{\text{qq}}^{\text{(1)}}[1,1,3,3]$ and $
   C_{\text{qq}}^{\text{(1)}}[1,3,3,1]$, is used to carry out the required analysis for the MFV scenario.\\

\begin{table}
\begin{center}
\begin{tabular}{|l|c|c|}
\hline
Measurement& Experiment& "Best" theory
\\
\hline\hline
$\Gamma_Z$(GeV) & $2.4955\pm 0.0023$ & $2.4943\pm 0.0006$   \cite{Freitas:2014hra,Dubovyk:2018rlg,Dubovyk:2019szj}  
\\
\hline
$R_e$ & $20.804\pm 0.05$ &  $20.732\pm 0.009 $ \cite{Freitas:2014hra,Dubovyk:2018rlg,Dubovyk:2019szj} \\
\hline
$R_\mu$ & $20.784\pm 0.034$ & $20.732\pm 0.009 $ \cite{Freitas:2014hra,Dubovyk:2018rlg,Dubovyk:2019szj}  \\
\hline 
$R_\tau$ & $20.764\pm 0.045$ & $ 20.779\pm 0.009$ \cite{Freitas:2014hra,Dubovyk:2018rlg,Dubovyk:2019szj}  \\
\hline 
$R_b$& $0.21629\pm 0.00066$ & $ 0.2159\pm 0.0001$\cite{Freitas:2014hra,Dubovyk:2018rlg,Dubovyk:2019szj}   \\
\hline
$R_c$ & $0.1721 \pm 0.0030$ & $ 0.1722\pm 0.00005 $\cite{Freitas:2014hra,Dubovyk:2018rlg,Dubovyk:2019szj}   \\
\hline
$\sigma_h$ & $41.481\pm 0.033 $ & $41.492\pm 0.008 $\cite{Freitas:2014hra,Dubovyk:2018rlg,Dubovyk:2019szj}    
\\
\hline 
$A_e ({\text{from}}~A_{LR}~{\text{had}}$ & $0.15138\pm 0.00216$ & $ 0.1469\pm 0.0004$ \cite{Dubovyk:2019szj,Awramik:2006uz}\\
\hline
$A_e ({\text{from}}~A_{LR}~{\text{lep}})$ & $0.1544\pm 0.0060$& $ 0.1469\pm 0.0004$ \cite{Dubovyk:2019szj,Awramik:2006uz}\\
\hline
$A_e ({\text{from~Bhabba~pol}})$ & $0.1498\pm 0.0049$ & $ 0.1469\pm 0.0004$ \cite{Dubovyk:2019szj,Awramik:2006uz}\\
\hline
$A_\mu$ & $0.142\pm 0.015$ & $ 0.1469\pm 0.0004$ \cite{Dubovyk:2019szj,Awramik:2006uz}\\
\hline
$A_\tau ({\text{from~SLD}})$ & $0.136\pm 0.015$ & $ 0.1469\pm 0.0004$ \cite{Dubovyk:2019szj,Awramik:2006uz}\\
\hline
$A_\tau (\tau~ {\text{pol}})$ & $0.1439\pm 0.0043$ & $ 0.1469\pm 0.0004$ \cite{Dubovyk:2019szj,Awramik:2006uz}\\
\hline
$A_c$ & $0.670\pm 0.027$ & $ 0.66773\pm 0.0002$\cite{Dubovyk:2019szj,Awramik:2006uz}\\
\hline 
$A_b$ & $0.923\pm 0.020$ & $0.92694\pm 0.00006$\cite{Dubovyk:2019szj,Awramik:2006uz,Awramik:2008gi}
\\ \hline  
$A_s$ & $0.895\pm 0.091$ & $  0.93563\pm 0.00004$\cite{Dubovyk:2019szj,Awramik:2006uz}\\
 \hline  
$A_{e,FB}$ & $0.0145\pm 0.0025$ & $ 0.0162\pm 0.0001$ \cite{Dubovyk:2019szj,Awramik:2006uz} \\
 \hline
$A_{\mu,FB}$ & $0.0169\pm 0.0013$ & $ 0.0162\pm 0.0001$ \cite{Dubovyk:2019szj,Awramik:2006uz} \\
 \hline
 $A_{\tau,FB}$ & $0.0188\pm 0.0017$ & $ 0.0162\pm 0.0001$ \cite{Dubovyk:2019szj,Awramik:2006uz} \\
 \hline
$A_{b,FB}$ & $0.0996\pm 0.0016$ & $ 0.1021\pm 0.0003$ \cite{Dubovyk:2019szj,Awramik:2006uz,Awramik:2008gi} \\
\hline
$A_{c,FB}$ & $0.0707\pm 0.0035$ & $ 0.0736\pm 0.0003$  \cite{Dubovyk:2019szj,Awramik:2006uz}\\
\hline
$A_{s,FB}$ & $0.0976\pm 0.0114$ & $0.10308 \pm 0.0003$ \cite{Dubovyk:2019szj,Awramik:2006uz} \\
\hline
$M_W $(GeV)~{\text{PDG~World~Ave}} & $80.377\pm 0.012$ &$ 80.357\pm 0.006$\cite{Awramik:2003rn,Erler:2019hds}\\ \hline
$\Gamma_W$(GeV)  & $2.085\pm 0.042$& $ 2.0903\pm  0.0003$\cite{Cho:2011rk}\\
\hline
\end{tabular}
\caption{Unless otherwise cited, experimental results are taken from Table 10.5 of the Particle Data Group\cite{PhysRevD.98.030001}.
The theory results include the full set of $2$-loop contributions
for the $Z$ pole observables, along with higher order corrections when known.     The theory predictions are
computed using the formulae in the indicated references and our input parameters, and the theory errors include
the parametric uncertainties on $M_t$ and $M_h$\cite{Dubovyk:2019szj}, along with the estimated theory 
uncertainties described in the respective papers.\label{tab:expnums}}
\end{center}
\end{table}

\begin{figure}[t]
	\centering
\includegraphics[width=4in]{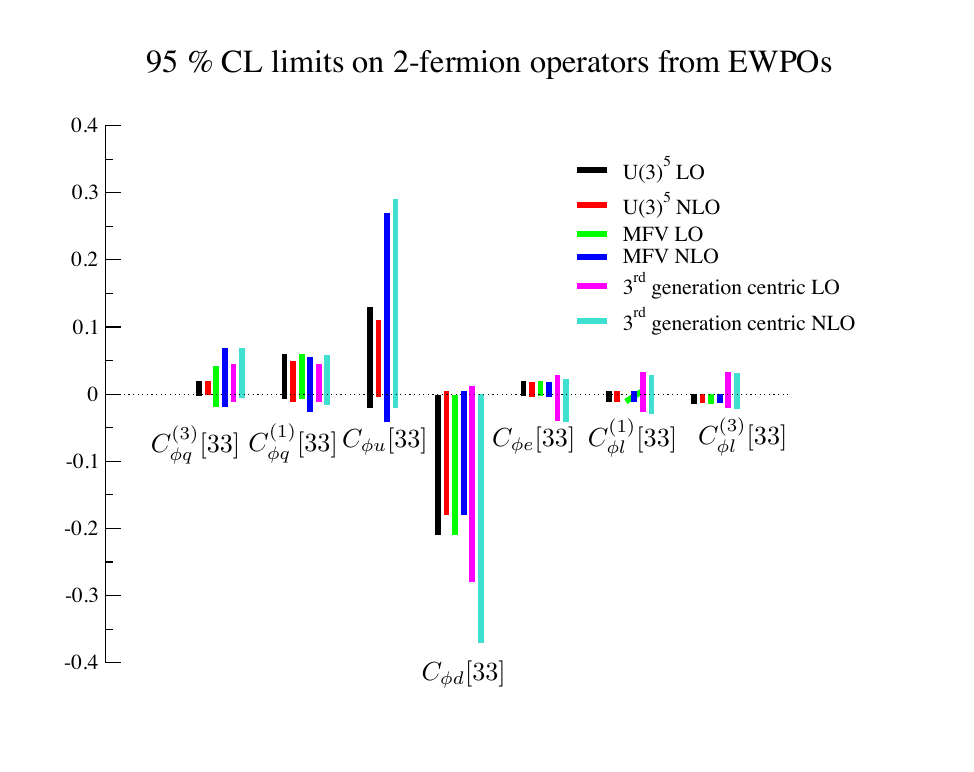}  
	\caption{Limits on coefficients of 2-fermion operators in the $U(3)^5$ (LO in black, NLO in red), MFV (LO in green, NLO in blue) and  $3^{rd}$ generation centric (LO in magenta, NLO in cyan) scenarios with a single non-zero operator and marginalizing over the other independent flavor structures of each operator. }
	\label{fig:fig1}
	\vskip .2in
\includegraphics[width=3in]{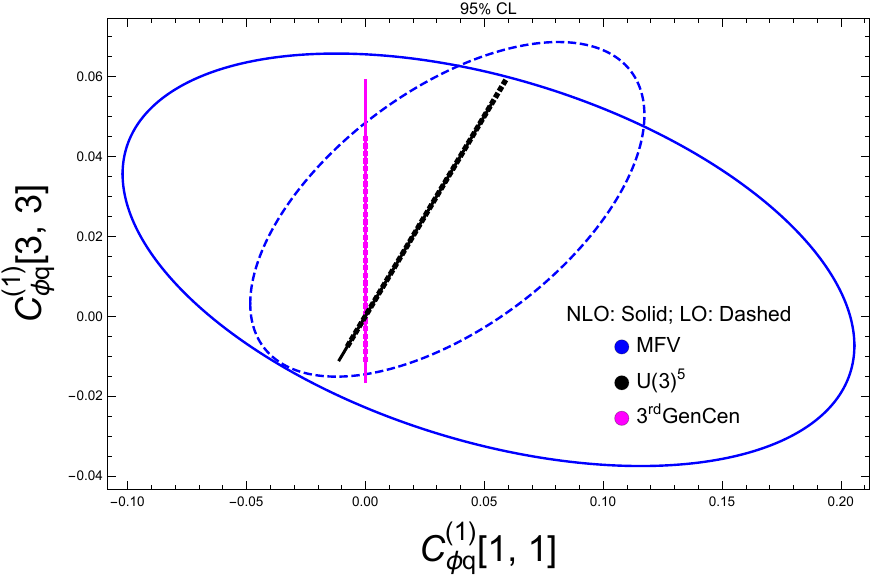}  \hskip .2in
\includegraphics[width=3in]{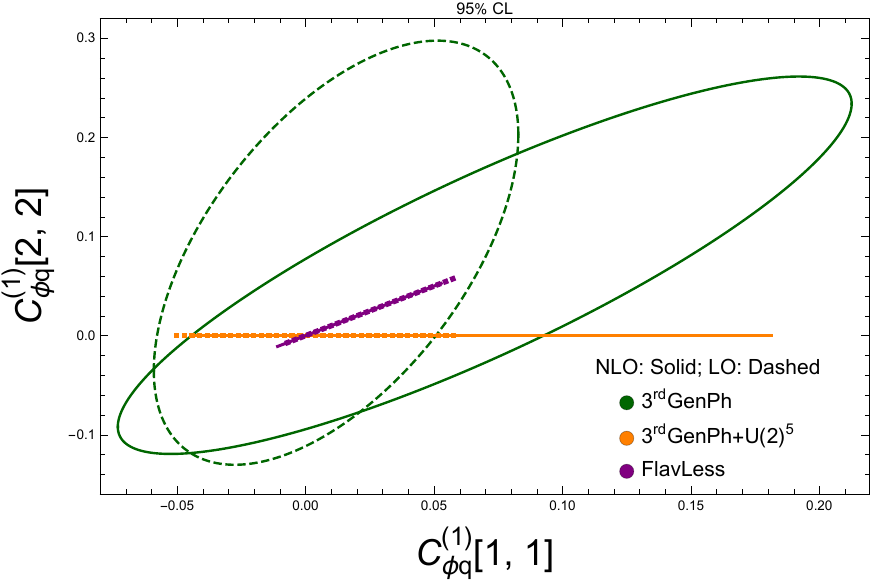} 
	\caption{95\% CL limits on $\C_{\phi q}^{(1)}[ij]$ under flavor assumptions described in the text. Results at LO are drawn with dashed lines, results at NLO are drawn with solid lines. {\bf On the left} we present $\C_{\phi q}^{(1)}[11]$ vs. $\C_{\phi q}^{(1)}[33]$ in the $U(3)^5$ (black), MFV (blue) and $3^{rd}$ generation centric (magenta) scenarios. In these scenarios $\C_{\phi q}^{(1)}[22]=\C_{\phi q}^{(1)}[11]$.  {\bf On the right} we present $\C_{\phi q}^{(1)}[11]$ vs. $\C_{\phi q}^{(1)}[22]$ in the $3^{rd}$ generation phobic (green), $3^{rd}$ generation phobic + $U(2)^5$ (orange) and flavorless (violet) scenarios. In the first two scenarios $\C_{\phi q}^{(1)}[33]=0$, while in the flavorless scenario $\C_{\phi q}^{(1)}[33]=\C_{\phi q}^{(1)}[22]=\C_{\phi q}^{(1)}[11]$. All other coefficients are set to 0.}
	\label{fig:hq1}
\end{figure}

We begin by considering limits on the 2-fermion operators (Class A) that contribute to the $Z$ and $W$  boson observables listed in Eq. \ref{eq:quan}.    Table \ref{tab:twoferm} compares
the LO and NLO results for the SMEFT coefficients in the $U(3)^5$, MFV, and $3^{rd}$ generation centric scenarios. 
We note that the $U(3)^5$ and flavorless scenarios  are identical for the coefficients of the 2-fermion operators. The limits on flavor structures that are not listed in the 
table can be derived  using the results of Section \ref{sec:smeft}, although for clarity we list 
limits on some of the non-independent coefficients.
    The differences between
the LO and NLO fits 
are in general quite small.         The single parameter limits are compared graphically in Fig. \ref{fig:fig1} where we 
see differences up to factors of 2 between the various flavor assumptions. 

The contributions of top quark loops to the 2-fermion operators have been studied in \cite{Liu:2022vgo}.  Table 3 of this reference presents the $95\%$ CL single parameter limits on $C_{uW}[33]$, $C_{uB}[33]$, $C_{\phi u}[33]$ and $C_{\phi q}^{(-)}[33]\equiv {1\over 2} ( C_{\phi q}^{(1)}[33]-C_{\phi q}^{(3)}[33])$.  Our results agree with these to  $\sim 10-20\%$, which is consistent with the use of slightly different sets of data as input.

In Fig. \ref{fig:hq1} and in Figs. \ref{fig:hq3}-\ref{fig:fhl3} in the appendix, we show on the left hand side  the limits on the 2-fermion operators  in the $U(3)^5$,  $3^{rd}$ generation specific, and MFV scenarios .  On the right hand side of these plots we show the $3^{rd}$ generation phobic and $3^{rd}$ generation phobic + $U(2)^5$ scenarios, where the new physics only couples to the first and second generations, and the flavorless scenario.  It is apparent that the limits one quotes on these operators is highly dependent on the assumed flavor scenario.  It is interesting to note in Fig. \ref{fig:hq1}, the large differences in the shapes of the LO and NLO fits in the MFV and $3^{rd}$ generation phobic scenarios and that the limits on the $3^{rd}$ generation phobic scenario are considerably weaker than in the other scenarios.

Tables \ref{tab:4flavB} and \ref{tab:4flavC} show the $95\%$ CL limits on coefficients in the  flavor schemes discussed in Sec. \ref{sec:smeft} for the Class B and Class C operators.  In general, there is a strong dependence on the flavor scenario.   This dependence is much larger than for the 2-fermion operator coefficients and the flavorless scenario gives much more stringent bounds for many coefficient functions than is the case in the other flavor scenarios. 

\begin{figure}[t]
	\centering
\includegraphics[width=3.5in]{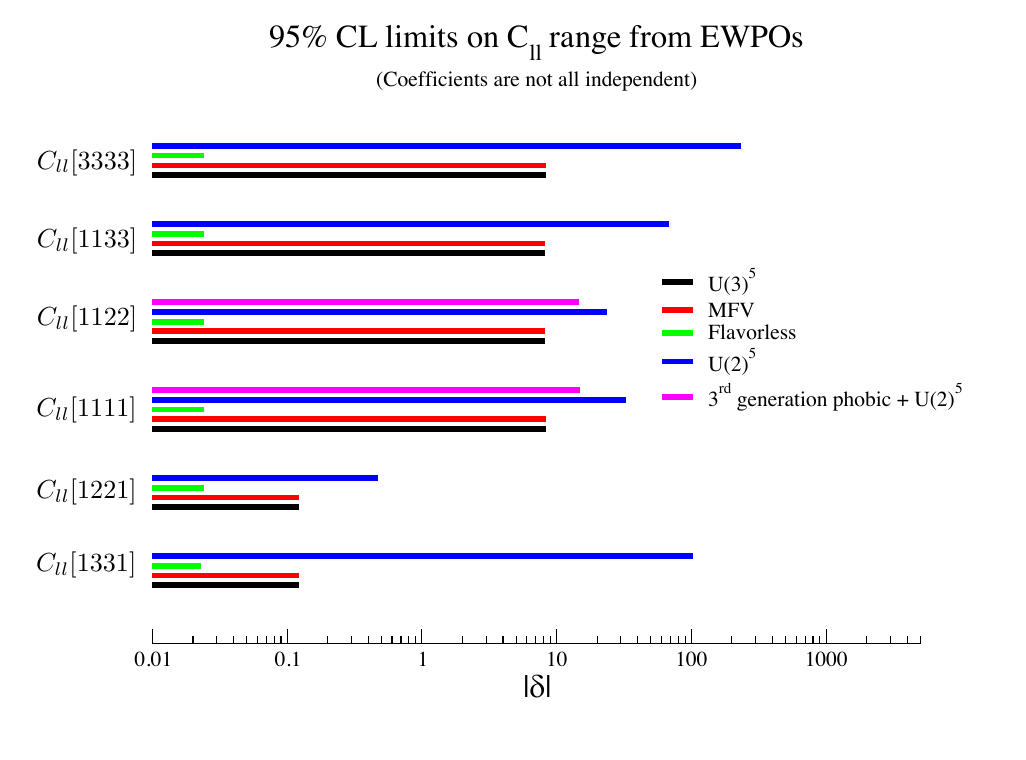} 
	\caption{Limits on coefficients of $\C_{ll}[ijkl]$ in various flavor scenarios, where $i,j,k,l=1,2,3$. Only $\C_{ll}$ is taken to be non-zero in this figure, and the independent flavor structures not shown are marginalized over.  Exact numbers are given in Table 
	\ref{tab:4flavB}. }
	\label{fig:figcll}
	\vskip .2in
\includegraphics[width=2.5in]{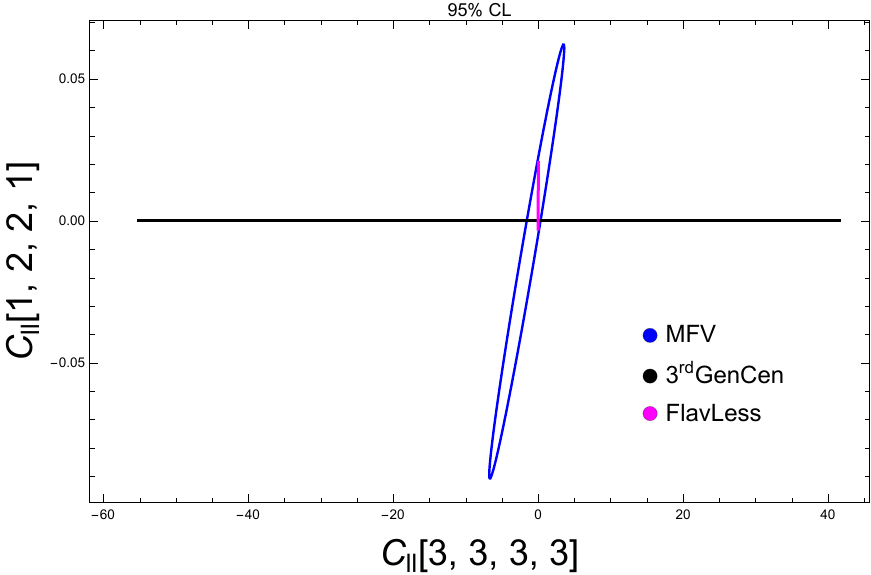} 
\hskip .2in
\includegraphics[width=2.5in]{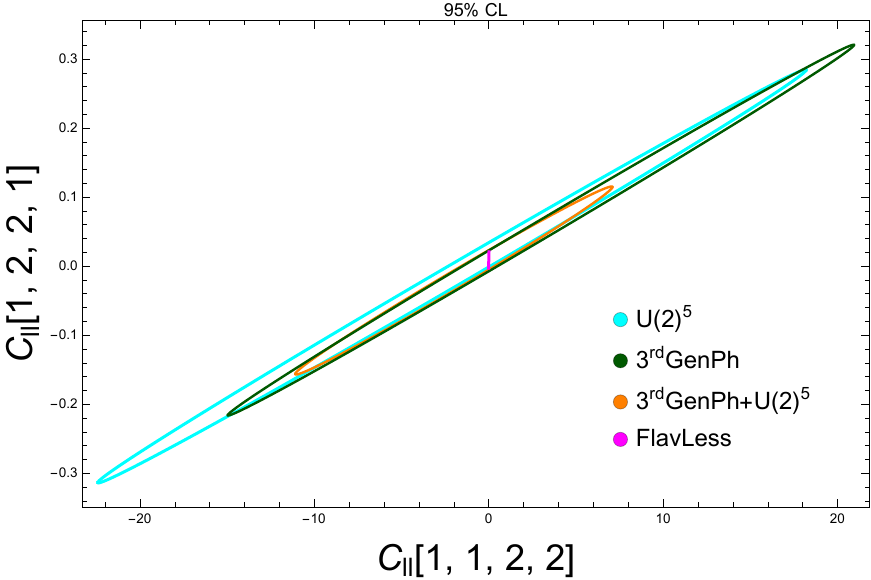} 
	\caption{95\% CL limits on $C_{ll}$
	 in various flavor scenarios with all other coefficients taken to be 0. {\bf On the left} we present $\C_{ll}[3333]$ vs. $\C_{ll}[1221]$ in the MFV (blue), $3^{rd}$ generation centric (black) and flavorless (magenta) scenarios. {\bf On the right} we present $\C_{ll}[1122]$ vs. $\C_{ll}[1221]$ in the $U(2)^5$ (cyan), $3^{rd}$ generation phobic (green), $3^{rd}$ generation phobic + $U(2)^5$ (orange) and flavorless (magenta) scenarios. All the independent flavor structures not present in the plots are marginalized over.}
	\label{fig:tll}
\end{figure}

The most precise limits 
are on $\C_{ll}$ and are shown in Fig. \ref{fig:figcll} for several flavor scenarios.  It is clear that using the flavorless scenario gives limits that are factors of ${\cal{O}}(10-100)$ more precise than in the MFV or $U(3)^5$ scenarios for $\C_{ll}$. This is understood from Eq. \ref{eq:gdef} where we see that 
the only flavor structure that contributes to $G_\mu$ is $\C_{ll}[1221]$ and we observe that the limits 
on $\C_{ll}$ in the $U(2)^5$
scenario are the weakest. 
In Fig. \ref{fig:tll}, we show the strong correlation between $\C_{ll}[1221]$ and $\C_{ll}[3333]$  in the MFV scenario and between 
 $\C_{ll}[1221]$ and $\C_{ll}[1122]$ in the $U(2)^5$ and  $3^{rd}$ generation phobic scenarios.  We note that there are only 2 independent $\C_{ll}$ 
 coefficients. 

\begin{figure}[t]
	\centering
        \includegraphics[width=3.5in]{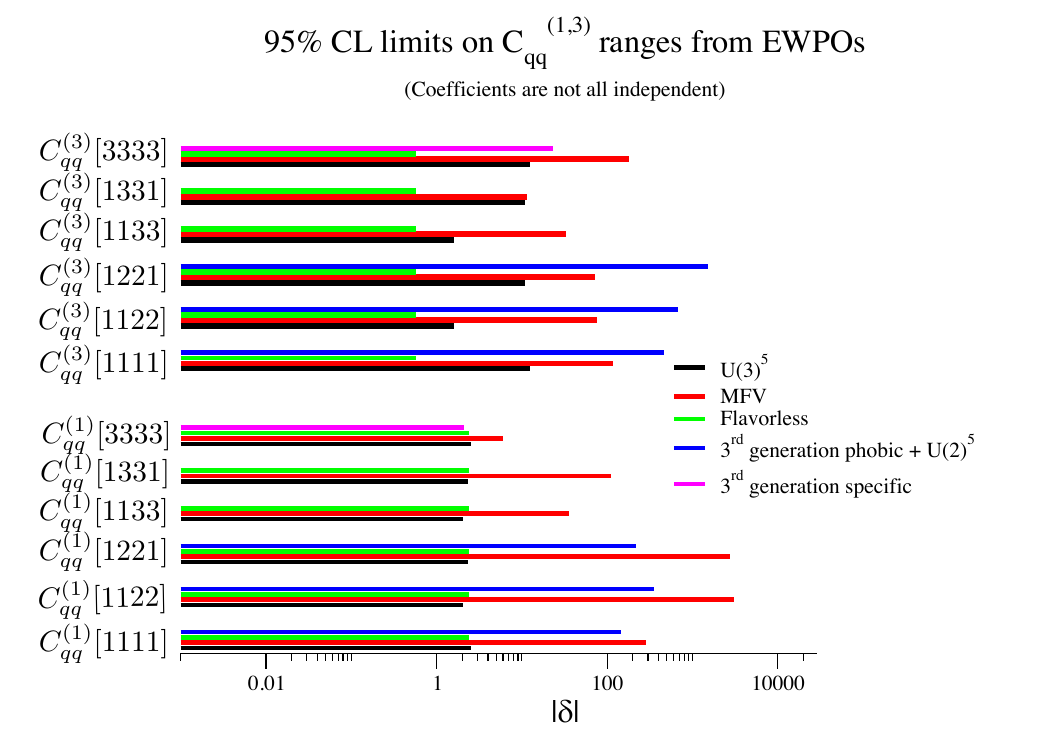} 
	\caption{Limits on sample of coefficients of Class B 4-fermion operators in various flavor scenarios. $\delta$ is
	the range of the $95\%$ confidence level limits.  Only a single operator is taken to be non-zero and the flavor structures not shown are marginalized over. Exact numbers are given in Table 
	\ref{tab:4flavB}. }
	\label{fig:figlu}
\vskip .2in
\includegraphics[width=2.5in]{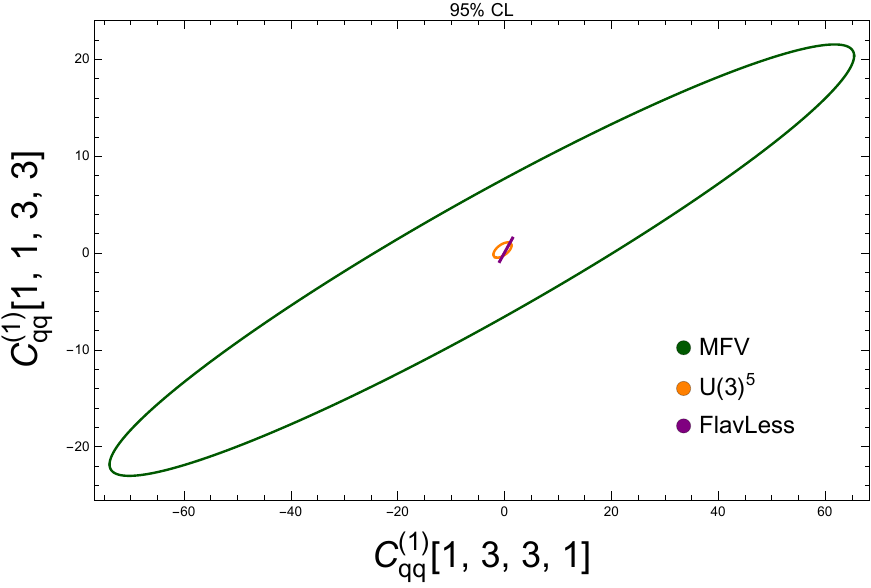} 
\hskip .2in
\includegraphics[width=2.5in]{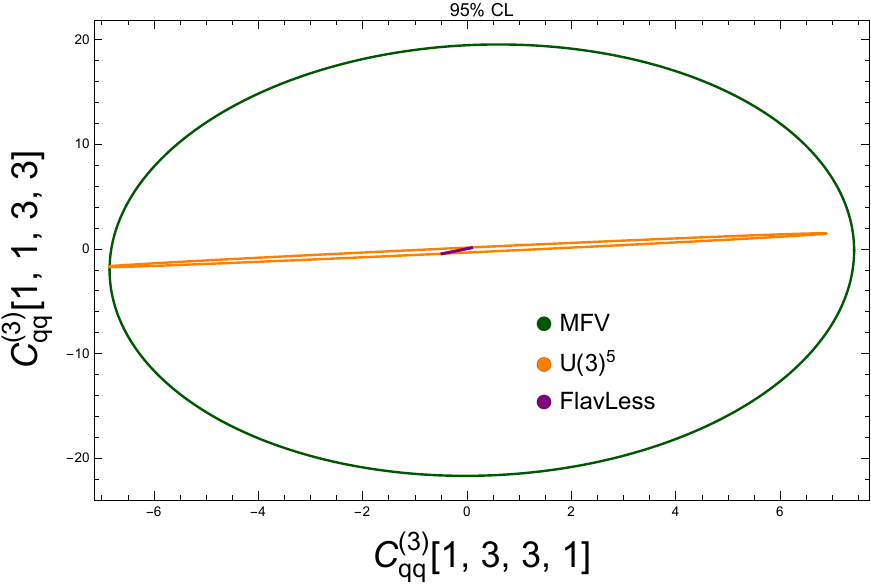} 
	\caption{95\% CL limits on $C_{qq}^{(1)}$ (left) and $C_{qq}^{(3)}$ (right), in various flavor scenarios with all other coefficients taken to be 0. {\bf On the left} we present $\C_{qq}^{(1)}[1331]$ vs. $\C_{qq}^{(1)}[1133]$ in the MFV (green), $3^{rd}$ generation centric (orange) and flavorless (violet) scenarios. {\bf On the right} we present $\C_{qq}^{(3)}[1331]$ vs. $\C_{qq}^{(3)}[1133]$ in the MFV (green), $3^{rd}$ generation centric (orange) and flavorless (violet) scenarios. All the independent flavor structures not present in the plots are marginalized over.}	\label{fig:tqq}
\end{figure}

Marginalized single parameter limits on the Class B operator coefficients $\C_{qq}^{(1)}$
and $\C_{qq}^{(3)}$ are shown in Fig. \ref{fig:figlu}. The $U(3)^5$ and MFV results for these
operators are within a factor of 2 of each other, while the $3^{rd}$ generation specific and $3^{rd}$
generation phobic scenarios are weakly constrained.  Sample correlations between different
flavor structures are shown in Fig. \ref{fig:tqq}, demonstrating the sensitivity to the  flavor assumptions of these fits.

\begin{figure}[t]
	\centering
        \includegraphics[width=3.5in]{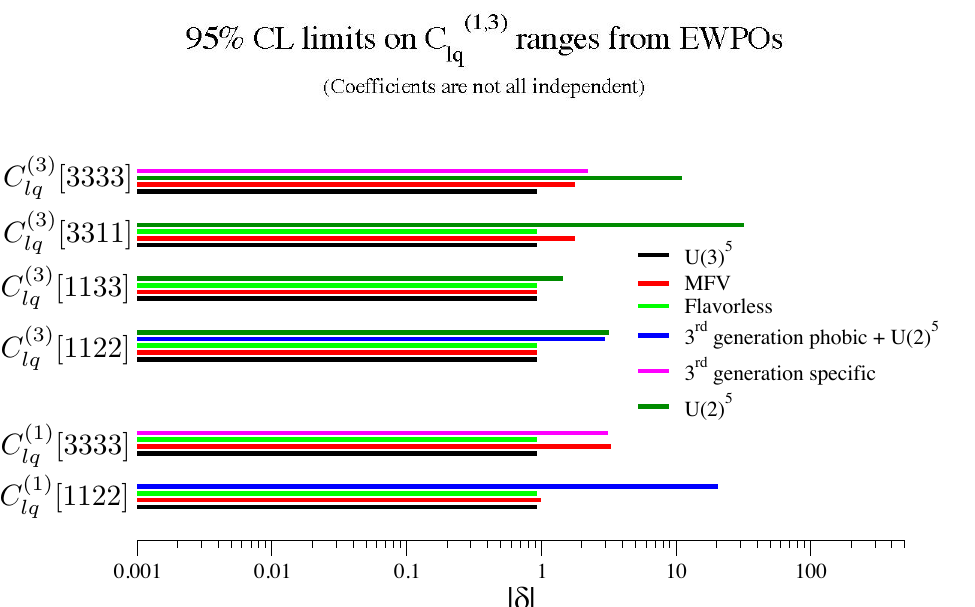} 
	\caption{Limits on sample of coefficients of Class C 4-fermion operators in various flavor scenarios, with the independent flavor structures of each operator marginalized over.  Exact numbers are given in \ref{tab:4flavC}. }
	\label{fig:lq3}
\vskip .2in
\includegraphics[width=2.5in]{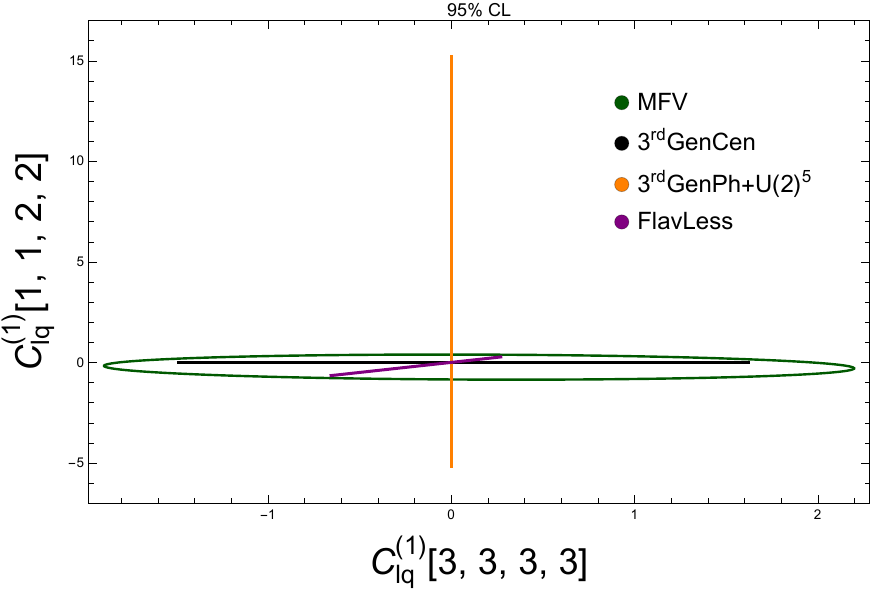} 
\hskip .2in
\includegraphics[width=2.5in]{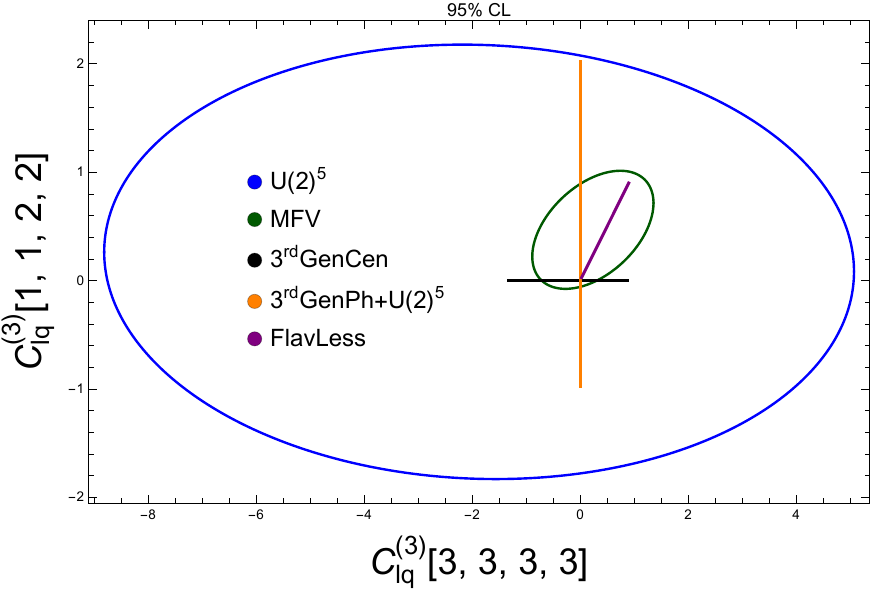} 
	\caption{95\% CL limits on $C_{lq}^{(1)}$ (left) and $C_{lq}^{(3)}$ (right), in various flavor scenarios with all other coefficients taken to be 0. {\bf On the left} we present $\C_{lq}^{(1)}[3333]$ vs. $\C_{lq}^{(1)}[1122]$ in the MFV (green), $3^{rd}$ generation centric (black), $3^{rd}$ generation phobic + $U(2)^5$ (orange) and flavorless (violet) scenarios. {\bf On the right} we present $\C_{lq}^{(3)}[3333]$ vs. $\C_{lq}^{(3)}[1122]$ in the $U(2)^5$ (blue), MFV (green), $3^{rd}$ generation centric (black), $3^{rd}$ generation phobic + $U(2)^5$ (orange) and flavorless (violet) scenarios.. All the independent flavor structures not present in the plots are marginalized over.}
	\label{fig:tlq}
\end{figure}

The 4-fermion Class C operators can be studied individually, and marginalizing over the independent flavor structures gives the limits  in Fig. \ref{fig:lq3}.  The limits obtained in the flavorless setup in this case also similar to those of the MFV and $U(3)^5$ flavor structures. The $3^{rd}$ generation specific and $3^{rd}$ generation phobic scenarios give weak bounds for these
operators.  In addition, there are large correlations between the different flavor structures, as illustrated in Fig. \ref{fig:tlq}.

\begin{table}
\begin{center}
\begin{tabular}{|l|c|c|c|c|c|c|}
\hline
Operator & LO $U(3)^5$ &NLO $U(3)^5$& LO MFV  & NLO  MFV & $3^{rd}$ gen centric & $3^{rd}$ gen centric
\\
&&&&&LO&NLO\\
\hline\hline
$\C_{\phi q}^{(3)}[33]$ &[-0.0029,0.020] &[-0.0024,0.020]&[-0.019,0.042]&[-0.019,0.069]&[-0.011,0.045]&[-0.0062,0.069]\\
$\C_{\phi q}^{(3)}[11]$ &[-0.0029,0.020]* &[-0.0024,0.020]*&[-0.0093,0.024]&[-0.012,0.020]&0&0\\
\hline
$\C_{\phi q}^{(1)}[33]$ &[-0.0070,0.060]&[-0.011,0.053]&[-0.0067,0.060]&[-0.027,0.055]&[-0.011,0.045]&[-0.016,0.059]\\ 
$\C_{\phi q}^{(1)}[11]$ &[-0.0070,0.060]*&[-0.011,0.053]*&[-0.032,0.10]&[-0.071,0.18]&0&0\\ 
\hline
$\C_{\phi u}[33]$&[-0.021,0.13]* & [-0.0037,0.11]&-&[-0.042,0.27]&-&[-0.020,0.29]\\
$\C_{\phi u}[11]$&[-0.021,0.13] &[-0.0037,0.11]*&[-0.021,0.13]&[-0.031,0.10]&0&0\\
\hline
$\C_{\phi d}[33]$ &[-0.21,-0.0024] &[-0.18,0.0036]&[-0.21,-0.0024]&[-0.18,0.0036]&[-0.28,0.012]&[-0.37,0.00083]\\
 \hline
 $\C_{\phi e}[33]$ &[-0.0033,0.019]&[-0.0036,0.018]&[-0.0033,0.019]&[-0.0036,0.018]&[-0.040,0.028]&[-0.041,0.022]\\
    \hline 
    $\C_{\phi l}^{(1)}[33]$ & [-0.011,0.0050]&{[-0.012,0.0049]}&[-0.011,0.0050]&[-0.012,0.0049]&[-0.026,0.033]&[-0.029,0.029]\\
     \hline
     $\C_{\phi l}^{(3)}[33]$ & [-0.015,-0.00041]&{[-0.013,-0.00029]}&[-0.015,-0.00041]&[-0.013,-0.00029]&[-0.020,0.033]&[-0.022,0.031]\\
     \hline
     $\C_{uW}[33]$&0&0&-&[-0.90,0.15]&-&[-0.90,0.15]\\
     \hline
     $\C_{uB}[33]$&0&0&-&[-0.61,0.061]&-&[-0.61,0.061]\\
     \hline\hline
\end{tabular}
\caption{95$\%$ CL allowed ranges on 2-fermion operators with varying flavor assumptions described in the text. Other flavor structure of a given operator are marginalized over, and different coefficients set to 0. The entries labelled with a "-" correspond to operators that do not contribute, while those labelled with a "*" are not independent. The flavor structures given 
not  in the table can be always derived from those in the table using the relations detailed in the text.\label{tab:twoferm} }
\end{center}
\end{table}

\begin{table}[t] 
\centering
\renewcommand{\arraystretch}{1.5}
\begin{tabular}{||c||c|c|c|c|c|c|c|||} 
\hline \hline
Operator & $U(3)^5$ & MFV &$U(2)^5$ &$3^{rd}$ gen  & $ 3^{rd}$ gen  & $3^{rd}$ gen  &Flavorless\\
&&&&specific&phobic & phobic+$U(2)^5$&
\\
\hline\hline 
$\C_{qq}^{(3)}[1133]$ &{[-0.80,0.81]}&[-17.6,15.4]&x&0&0&0&[-0.48,0.09]\\
$\C_{qq}^{(3)}[1331]$ &{[-4.38,6.42]}&[-5.43,6.00]&x&0&0&0&[-0.48,0.09]*\\
$\C_{qq}^{(3)}[1122]$ &{[-0.80,0.81]*}&{[-31.2,45.4]}&x&0& x&[-303,375]&[-0.48,0.09]*\\
$\C_{qq}^{(3)}[1221]$ &{[-4.38,6.42]*}&[-44.8,26.3]&x&0& x&[-836,672]&[-0.48,0.09]*\\
$\C_{qq}^{(3)}[3333]$ &{[-5.15,7.21]*}&{[-89.4,90.5]*}&x&[-2.94,20.4]&0&0&[-0.48,0.09]*\\
$\C_{qq}^{(3)}[1111]$ &{[-5.15,7.21]*}&[-60.9,56.6]*&x&0&x&{[-460,369]*}&[-0.48,0.09]*\\
\hline\hline
$\C_{qq}^{(1)}[1133]$ &{[-0.20,1.84]}&[-18.6,17.1]&x&0&0&0&[-0.89,1.52]\\
$\C_{qq}^{(1)}[1331]$ &{[-1.39,0.94]}&[-60.0,51.6]&x&0&0&0&[-0.89,1.52]*\\
$\C_{qq}^{(1)}[1122]$ &{[-0.20,1.84]*}&{[-1623,1408]}&x&0&x&[-189,169]&[-0.89,1.52]*\\
$\C_{qq}^{(1)}[1221]$ &{[-1.39,0.94]*}&[-1276,1470]&x&0&x&[-109,106]&[-0.89,1.52]*\\
$\C_{qq}^{(1)}[3333]$ &{[-0.66,1.86]*}&{[-2.62,3.42]*}&x&[-0.26,1.84]&0&0&[-0.89,1.52]*\\
$\C_{qq}^{(1)}[1111]$ &{[-0.66,1.86]*}&[-153,133]*&x&0&x&{[-85.0,62.3]*}&[-0.89,1.52]*\\
\hline\hline
$\C_{ll}[1133]$ &{[-5.65,2.52]}&[-5.65,2.52]&[-46.1,22.2]&0&0&0&[-.0029,.021]\\
$\C_{ll}[1331]$ &{[-0.076,0.047]}&[-0.076,0.047]&[-38.2,65.4]&0&0&0&[-.0029,.021]*\\
$\C_{ll}[1122]$ &{[-5.65,2.52]*}&[-5.65,2.52]*&[-18.4,14.2]&0&[-11.4,17.4]&[-9.3,5.3]&[-.0029,.021]*\\
$\C_{ll}[1221]$ &{[-0.076,0.047]*}&[-0.076,0.047]*&[-0.25,0.22]&0&[-0.16,0.27]&[-0.13,0.088]&[-.0029,.021]*\\
$\C_{ll}[3333]$&{[-5.72,2.56]*}&[-5.72,2.56]*&[-44.0,191]&[-55.2,41.6]&0&0&[-.0029,.021]*\\
$\C_{ll}[2222]$ &{[-5.72,2.56]*}&[-5.72,2.56]*&[-18.6,14.4]*&0&[-90.2,27.0]&[-9.4,5.4]*&[-.0029,.021]*\\
$\C_{ll}[1111]$ &{[-5.72,2.56]*}&[-5.72,2.56]*&[-18.6,14.4]*&0&[-26.7,26.3]&[-9.4,5.4]*&[-.0029,.021]*\\
\hline\hline
$\C_{ee}[1133]$ &[-0.80,4.07]&[-0.80,4.07]&x&0&0&0&[-1.07,5.4]\\
$\C_{ee}[1122]$ &[-0.80,4.07]*&[-0.80,4.07]*&x&0&x&[-0.67,6.02]&[-1.07,5.4]*\\
$\C_{ee}[3333]$ &[-1.60,8.13]*&[-1.60,8.13]*&x&[-36.7,19.7]&0&0&[-1.07,5.4]*\\
$\C_{ee}[1111]$ &[-0.80,4.07]*&[-0.80,4.07]*&x&0&x&[-1.34,12.0]&[-1.07,5.4]*\\
\hline\hline
$\C_{uu}[1133]$ &x&x&x&x&x&x&[-1.67,0.30]\\
\hline\hline
$\C_{dd}[3333]$ &x&x&x&[-428,5.54]&x&x&[-68.4,2.38]\\
\hline\hline
\end{tabular}
\caption{NLO results for $95\%$ CL limits on Class B operators when the coefficients of  all other operators are set to 0 and
the different flavor structures of a given operator are marginalized over.  The limits marked with a "*" are obtained using the relations given in the text and do not represent independent coefficients.  We label with an "x" the cases where no limit can be derived.\label{tab:4flavB}}
\end{table}

\begin{table}[t] 
\centering
\renewcommand{\arraystretch}{1.5}
\begin{tabular}{||c||c|c|c|c|c|c||} 
\hline \hline
Operator & $U(3)^5$ & MFV &$U(2)^5$ &$3^{rd}$ gen specific & $ 3^{rd}$ gen phobic +$U(2)^5$& Flavorless
\\
\hline\hline 
$\C_{lq}^{(3)}[1133]$ &[0.026,0.90] &[0.03,0.90]&[-0.0078,1.45]&0&0&[0.026,0.90]\\
$\C_{lq}^{(3)}[3311]$ &[0.026,0.90]*&[-0.66,1.13]&[-9.62,22.2]&0&0&[0.026,0.90]*\\
$\C_{lq}^{(3)}[1122]$ &[0.026,0.90]*&[0.03,0.90]*&[-1.43,1.78]&0&[-0.98,2.02]&[0.026,0.90]*\\
$\C_{lq}^{(3)}[3333]$ &[0.026,0.90]*&[-0.66,1.13]*&[-7.43,3.69]&[-1.33,0.88]&0&[0.026,0.90]*\\
\hline\hline
$\C_{lq}^{(1)}[1122]$ &[-0.66,0.27]&[-0.73,0.27]&x&0&[-5.20,15.2]&[-0.66,0.27]\\
$\C_{lq}^{(1)}[3333]$ &[-0.66,0.27]*&[-1.50,1.79]&x&[-1.50,1.62]&0&[-0.66,0.27]*\\
\hline\hline
$\C_{lu}[1122]$ &[-0.20,0.49]&[-0.20,0.55]&x&0&[-2.6,7.6]&[-0.20,0.49]\\
$\C_{lu}[3333]$ &[-0.20,0.49]*&[-1.33,1.13]&x&[-1.35,1.27]&0&[-0.20,0.49]*\\
\hline\hline
$\C_{qe}[1122]$ &[-0.20,1.06]&[-66.2,142]&x&0&[-25.7,3.34]&[-0.20,1.06]\\
$\C_{qe}[3333]$ &[-0.20,1.06]*&[-2.38,6.35]&x&[-2.26,1.25]&0&[-0.20,1.06]*\\
\hline\hline
$\C_{ed}[1122]$ &[-3.86,15.3]&[-3.86,15.3]&x&0&[-3.29,25.8]&[-3.86,15.3]\\
$\C_{ed}[3333]$ &[-3.86,15.3]*&[-3.86,15.3]*&x&[-117,42.8]&0&[-3.86,15.3]*\\
\hline\hline
$\C_{ld}[1122]$ &[-9.11,3.26]&[-9.11,3.26]&x&0&[-15.2,5.21]&[-9.11,3.26]\\
$\C_{ld}[3333]$ &[-9.11,3.26]*&[-9.11,3.26]*&x&{[-72.9,59.3]}&0&[-9.11,3.26]*\\
\hline\hline 
$\C_{le}[1122]$ &[-7.09,8.13]&[-7.09,8.13]&x&0&[-10.5,14.5]&[-7.09,8.13]\\
$\C_{le}[3333]$ &[-7.09,8.13]*&[-7.09,8.13]*&x&[-92.8,55.5]&0&[-7.09,8.13]*\\
\hline\hline
$\C_{eu}[1122]$&[-0.81,0.16]&[-0.88,0.11]&x&0&[-12.9,1.66]&[-0.81,0.16]\\
$\C_{eu}[3333]$&[-0.81,0.16]*&[-1.07,1.73]&x&[-1.05,1.95]&0&[-0.81,0.16]*\\
\hline\hline
$\C_{ud}^{(1)}[1122]$&[-0.30,7.68]&[-154,90.2]&x&0&[-13.9,75.7]&[-0.30,7.68]\\
$\C_{ud}^{(1)}[3333]$&[-0.30,7.68]*&[-9.3,27.0]&x&[-0.24,18.4]&0&[-0.30,7.68]*\\
\hline\hline
$\C_{qd}^{(1)}[1122]$&[-13.3,0.47]&[-202,128]&x&0&[-29.4,139]&[-13.3,0.47]\\
$\C_{qd}^{(1)}[3333]$&[-13.3,0.47]*&[-27.2,8.27]&x&[-24.4,0.11]&0&[-13.3,0.47]*\\
\hline\hline
$\C_{qu}^{(1)}[1122]$&[-2.82,0.54]&x&x&0&[-71.0,14.1]&[-2.82,0.54]\\
$\C_{qu}^{(1)}[3333]$&[-2.82,0.54]*&x&x&[-3.07,0.44]&0&[-2.82,0.54]*\\
\hline\hline
\end{tabular}
\caption{NLO results for $95\%$ CL limits on Class C operators when the coefficients of  all other operators are set to 0 and
the different flavor structures of a given operator are marginalized over.  The limits marked with a "*" are obtained using the relations given in the text and do not represent independent coefficients.  We label with an "x" the cases where no limit can be derived.\label{tab:4flavC}}
\end{table}

\section{Conclusions}
Measurements at the $Z$ and $W$ boson poles provide important limits on BSM physics.   Using tree level predictions, the
fermion-gauge boson interactions are restricted to be quite close to those of the SM.  In the context of the SMEFT,
the bounds on anomalous particle couplings can be extended to systematically include NLO QCD and EW effects and interpreted as bounds on dimension-6 SMEFT operators.  These bounds were previously derived ignoring
the effects of the flavor dependence of the dimension-6  2- and 4-fermion operators. 
We have re-computed the NLO contributions to EWPOs including an arbitrary flavor structure for the fermion operators.  
With the exception of $\C_{ll}$ (which contributes to muon decay and is quite precisely bounded), the operators that contribute to EWPOs at tree level
involve 2 fermions and the NLO corrections are typically quite small relative to the LO results. 
  Comparing the NLO bounds on 2-fermion operators in different flavor scenarios results in factors of ${\cal{O}}(2)$.

 However, the flavor structure has a dramatic effect on the bounds derived on the 4-fermion
operators which first contribute to EWPOs at NLO.  We present results for 
 4-fermion operators when the different
flavor structures of an operator are marginalized over and observe large deviations in the fits for different flavor 
scenarios.  These large effects result from the significant correlations between the different flavor structures.
 We have explicitly demonstrated that ignoring flavor typically leads to much tighter bounds on the SMEFT coefficients
 than in a general flavor scenario. 
 
 We also notice that the presence of a top quark in the internal propagators generally leads to stronger constraints. As we pointed out, the effects coming from the inclusion of NLO contributions are larger for operators where the top quark can contribute in the loop corrections. Furthermore, the limits that can be obtained on the 4-fermion operator coefficients in the $U(3)$ scenario are often orders of magnitude stronger than those obtained in the $3^{rd}$ generation phobic + $U(2)^5$ scenario. Since these scenarios have the same number of independent coefficients, the difference cannot be attributed to the marginalization procedure. Moreover, in the case of $\mathcal{O}_{ll}$, $\mathcal{O}_{ee}$ and $\mathcal{O}_{le}$, where we do not expect an enhancement due to top loops, the two scenarios produce bounds with similar order of magnitudes.  
 
  At tree level, many of these 4-fermion operators contribute to the well measured Drell Yan process\cite{deBlas:2013qqa,Panico:2021vav,Cirigliano:2012ab}.  The NLO and EW corrections to Drell Yan  production can potentially be combined with the EWPOs studied here to further 
 improve the precision of the limits on the SMEFT coefficients\cite{Dawson:2021ofa,Dawson:2018jlg,Boughezal:2023nhe,Boughezal:2022nof,Breso-Pla:2021qoe}. 
  
  Numerical results for the EWPOs to NLO in the QCD and EW interactions for a completely arbitrary flavor structure
  of the 2-fermion and 4-fermion operators are presented in the supplemental material.  The supplemental material also contains numerical results for $H\rightarrow \gamma\gamma$ and $h
  \rightarrow Z \gamma$ to NLO in the QCD and EW interactions for a completely arbitrary flavor structure
  of the 2-fermion and 4-fermion operators, generalizing the results of \cite{Dawson:2018pyl,Dawson:2018liq}.  In the case of the Higgs decays, the effects of the flavor structure are small. 
\label{sec:conc}
\section*{Acknowledgements} 
We thank A. Biekoetter,  B. Pecjak, D. Scott, and T. Smith for a detailed comparison of the NLO EW results
for the EWPOs. L.B. would like to thank A. Brossa Gonzalo for helpful discussions.
S.D.   is  supported by the U.S. Department of Energy under Grant Contract  de-sc0012704.  
The work of L.B. and P.P.G. has received financial support from Xunta de Galicia (Centro
singular de investigaci\'on de Galicia accreditation 2019-2022), by
European Union ERDF, and by ``Mar\'ia de Maeztu" Units of Excellence
program MDM-2016-0692 and the Spanish Research State Agency.

\bibliographystyle{utphys}
\bibliography{flavor_paper.bib}
\appendix
\section{Other plots for 2-fermion operators}
\label{sec:app}

In this appendix, we show a collection of plots involving the 2-fermion operators, as described in the text.

\begin{figure}[ht]
	\centering
\includegraphics[width=3in]{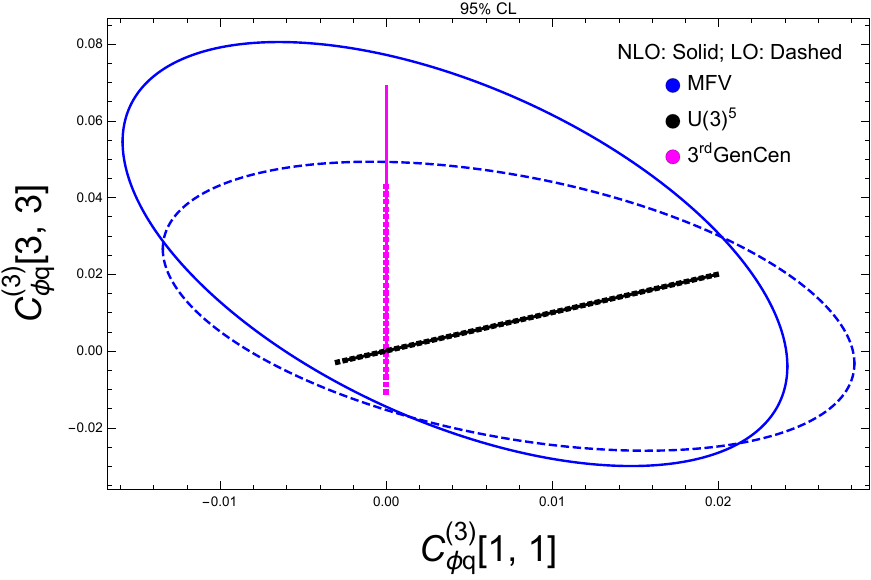}  \hskip .2in
\includegraphics[width=3in]{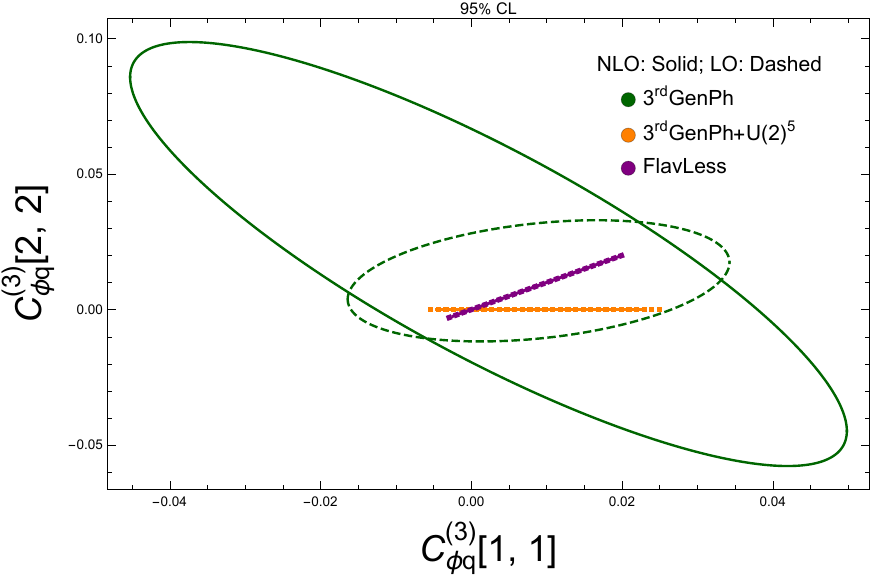} 
	\caption{95\% CL limits on $\C_{\phi q}^{(3)}[ij]$ under flavor assumptions described in the text. Results at LO are drawn with dashed lines, results at NLO are drawn with solid lines. {\bf On the left} we present $\C_{\phi q}^{(3)}[11]$ vs. $\C_{\phi q}^{(3)}[33]$ in the $U(3)^5$ (black), MFV (blue) and $3^{rd}$ generation centric (magenta) scenarios. In these scenarios $\C_{\phi q}^{(3)}[22]=\C_{\phi q}^{(3)}[11]$.  {\bf On the right} we present $\C_{\phi q}^{(3)}[11]$ vs. $\C_{\phi q}^{(3)}[22]$ in the $3^{rd}$ generation phobic (green), $3^{rd}$ generation phobic + $U(2)^5$ (orange) and flavorless (violet) scenarios. In the first two scenarios $\C_{\phi q}^{(3)}[33]=0$, while in the flavorless scenario $\C_{\phi q}^{(3)}[33]=\C_{\phi q}^{(3)}[22]=\C_{\phi q}^{(3)}[11]$. All other coefficients are set to 0.}
	\label{fig:hq3}
\end{figure}

\begin{figure}[ht]
	\centering
\includegraphics[width=3in]{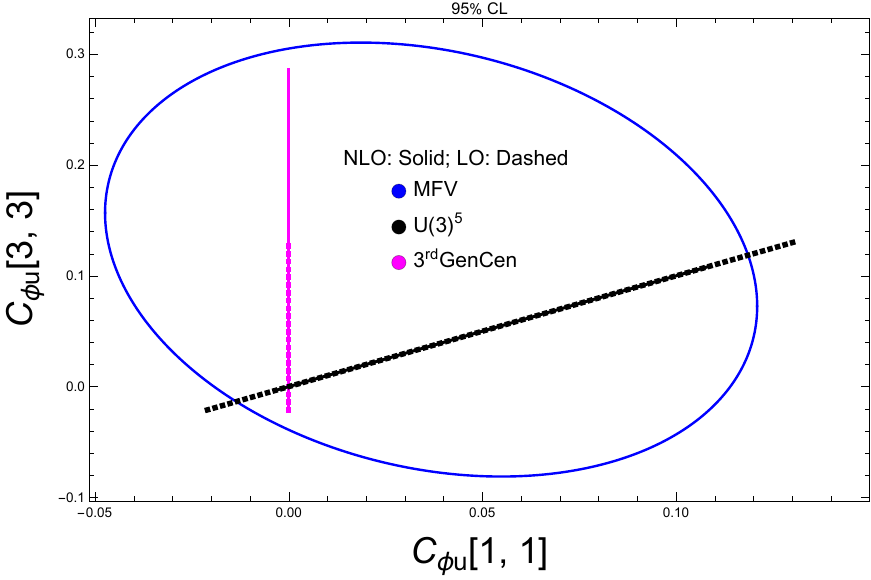}  \hskip .2in
\includegraphics[width=3in]{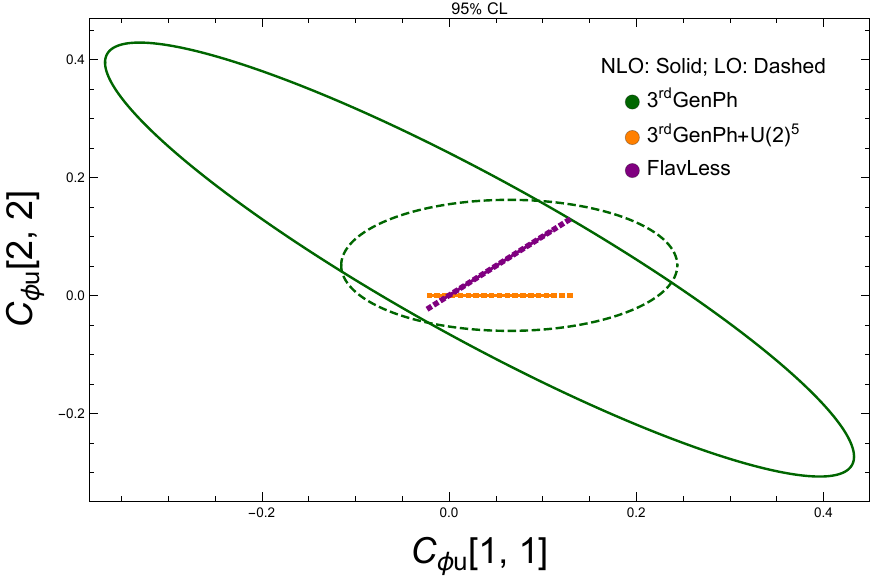} 
	\caption{95\% CL limits on $\C_{\phi u}[ij]$ under flavor assumptions described in the text. Results at LO are drawn with dashed lines, results at NLO are drawn with solid lines. {\bf On the left} we present $\C_{\phi u}[11]$ vs. $\C_{\phi u}[33]$ in the $U(3)^5$ (black), MFV (blue) and $3^{rd}$ generation centric (magenta) scenarios. In these scenarios $\C_{\phi u}[22]=\C_{\phi u}[11]$, also notice that for the MFV scenario only the NLO results can be obtained.  {\bf On the right} we present $\C_{\phi u}[11]$ vs. $\C_{\phi u}[22]$ in the $3^{rd}$ generation phobic (green), $3^{rd}$ generation phobic + $U(2)^5$ (orange) and flavorless (violet) scenarios. In the first two scenarios $\C_{\phi u}[33]=0$, while in the flavorless scenario $\C_{\phi u}[33]=\C_{\phi u}[22]=\C_{\phi u}[11]$. All other coefficients are set to 0.}
	\label{fig:fu}
\end{figure}

\begin{figure}[ht]
	\centering
\includegraphics[width=3in]{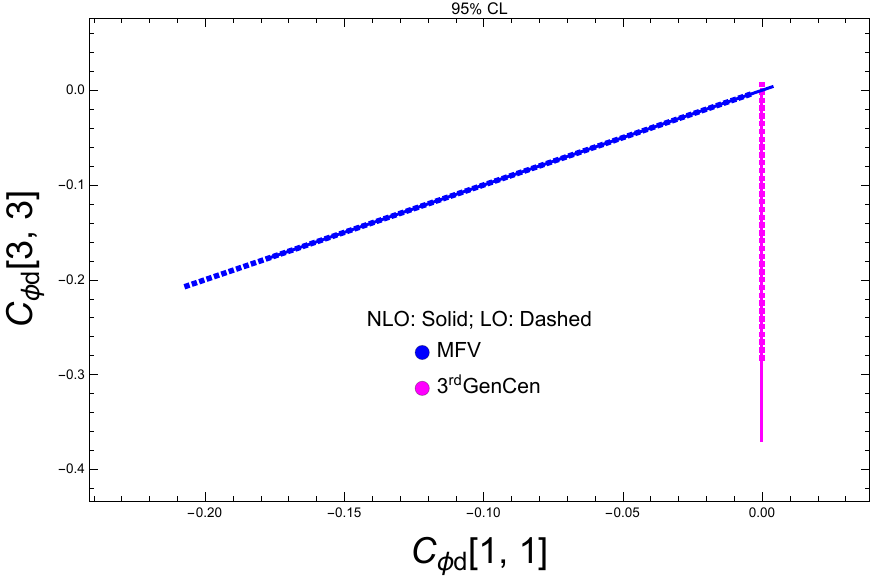}  \hskip .2in
\includegraphics[width=3in]{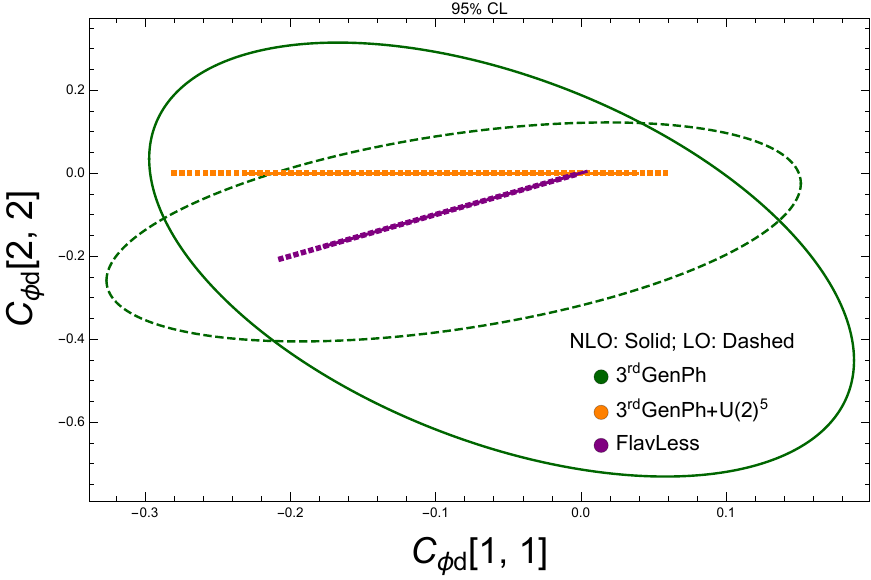} 
	\caption{95\% CL limits on $\C_{\phi d}[ij]$ under flavor assumptions described in the text. Results at LO are drawn with dashed lines, results at NLO are drawn with solid lines. {\bf On the left} we present $\C_{\phi d}[11]$ vs. $\C_{\phi d}[33]$ in the MFV (blue) and $3^{rd}$ generation centric (magenta) scenarios. In these scenarios $\C_{\phi d}[22]=\C_{\phi d}[11]$.  {\bf On the right} we present $\C_{\phi d}[11]$ vs. $\C_{\phi d}[22]$ in the $3^{rd}$ generation phobic (green), $3^{rd}$ generation phobic + $U(2)^5$ (orange) and flavorless (violet) scenarios. In the first two scenarios $\C_{\phi d}[33]=0$, while in the flavorless scenario $\C_{\phi d}[33]=\C_{\phi d}[22]=\C_{\phi d}[11]$. All other coefficients are set to 0.}
	\label{fig:fd}
\end{figure}

\begin{figure}[ht]
	\centering
\includegraphics[width=3in]{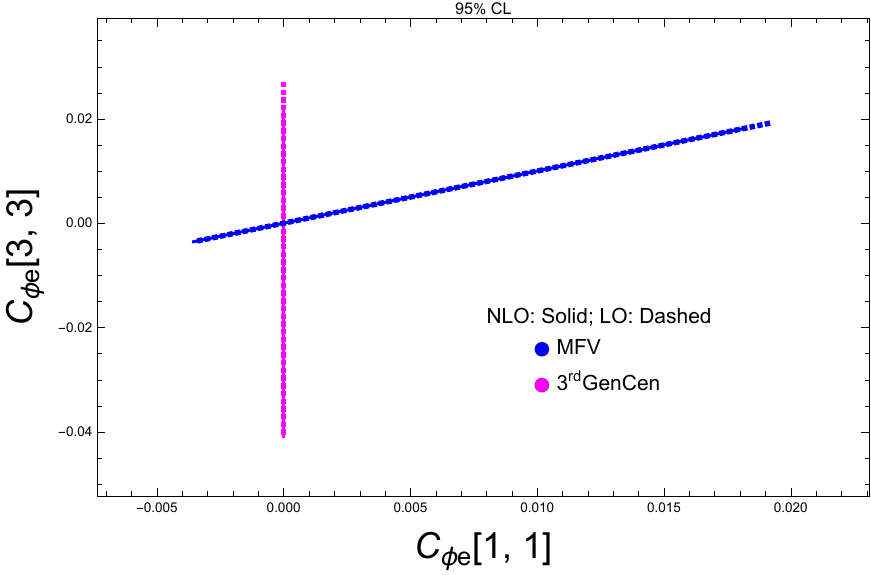}  \hskip .2in
\includegraphics[width=3in]{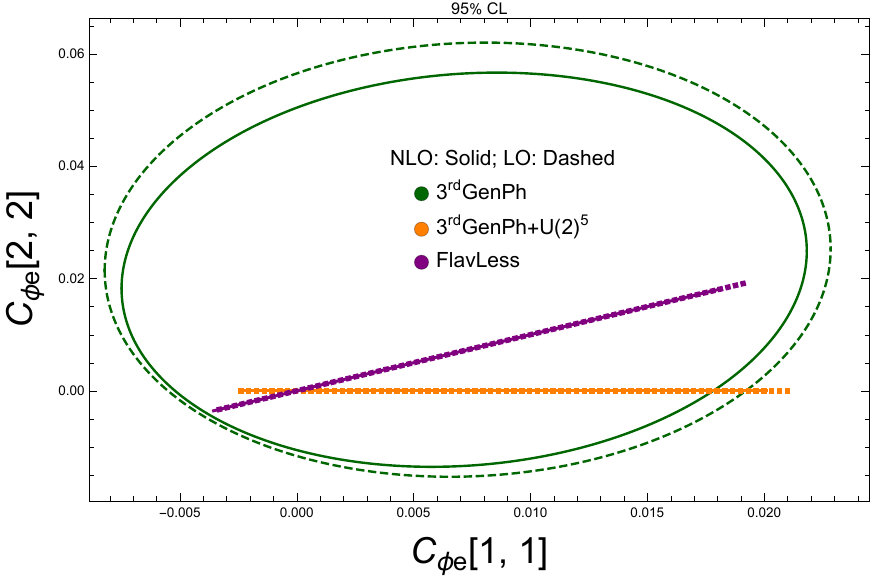} 
	\caption{95\% CL limits on $\C_{\phi e}[ij]$ under flavor assumptions described in the text. Results at LO are drawn with dashed lines, results at NLO are drawn with solid lines. {\bf On the left} we present $\C_{\phi e}[11]$ vs. $\C_{\phi e}[33]$ in the MFV (blue) and $3^{rd}$ generation centric (magenta) scenarios. In these scenarios $\C_{\phi e}[22]=\C_{\phi e}[11]$.  {\bf On the right} we present $\C_{\phi e}[11]$ vs. $\C_{\phi e}[22]$ in the $3^{rd}$ generation phobic (green), $3^{rd}$ generation phobic + $U(2)^5$ (orange) and flavorless (violet) scenarios. In the first two scenarios $\C_{\phi e}[33]=0$, while in the flavorless scenario $\C_{\phi e}[33]=\C_{\phi e}[22]=\C_{\phi e}[11]$. All other coefficients are set to 0.}
	\label{fig:fe}
\end{figure}

\begin{figure}[ht]
	\centering
\includegraphics[width=3in]{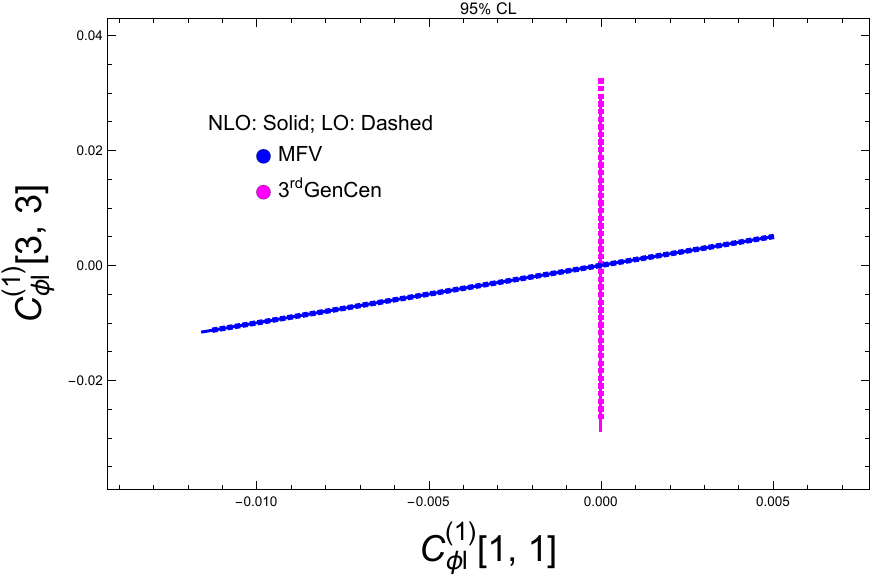}  \hskip .2in
\includegraphics[width=3in]{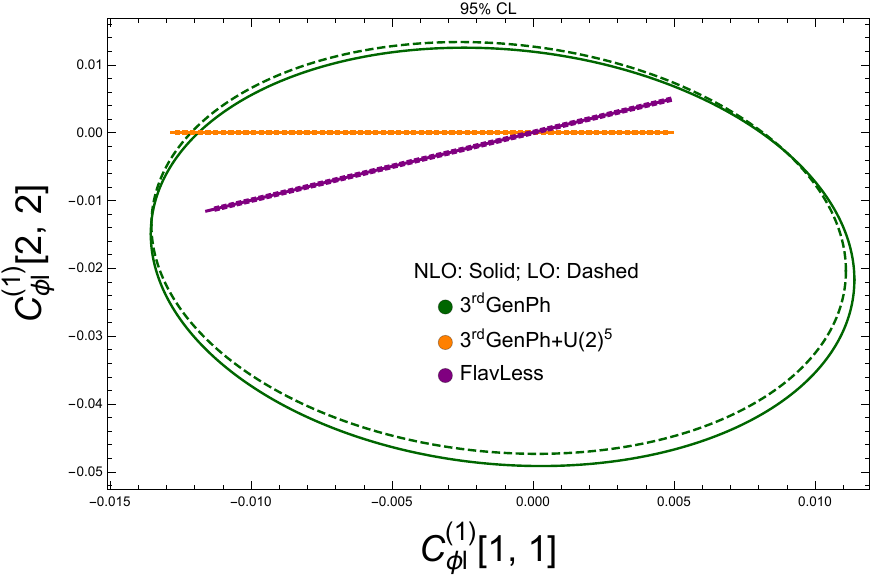} 
	\caption{95\% CL limits on $\C_{\phi l}^{(1)}[ij]$ under flavor assumptions described in the text. Results at LO are drawn with dashed lines, results at NLO are drawn with solid lines. {\bf On the left} we present $\C_{\phi l}^{(1)}[11]$ vs. $\C_{\phi l}^{(1)}[33]$ in the MFV (blue) and $3^{rd}$ generation centric (magenta) scenarios. In these scenarios $\C_{\phi l}^{(1)}[22]=\C_{\phi l}^{(1)}[11]$.  {\bf On the right} we present $\C_{\phi l}^{(1)}[11]$ vs. $\C_{\phi l}^{(1)}[22]$ in the $3^{rd}$ generation phobic (green), $3^{rd}$ generation phobic + $U(2)^5$ (orange) and flavorless (violet) scenarios. In the first two scenarios $\C_{\phi l}^{(1)}[33]=0$, while in the flavorless scenario $\C_{\phi l}^{(1)}[33]=\C_{\phi l}^{(1)}[22]=\C_{\phi l}^{(1)}[11]$. All other coefficients are set to 0.}
	\label{fig:fhl1}
\end{figure}

\begin{figure}[ht]
	\centering
\includegraphics[width=3in]{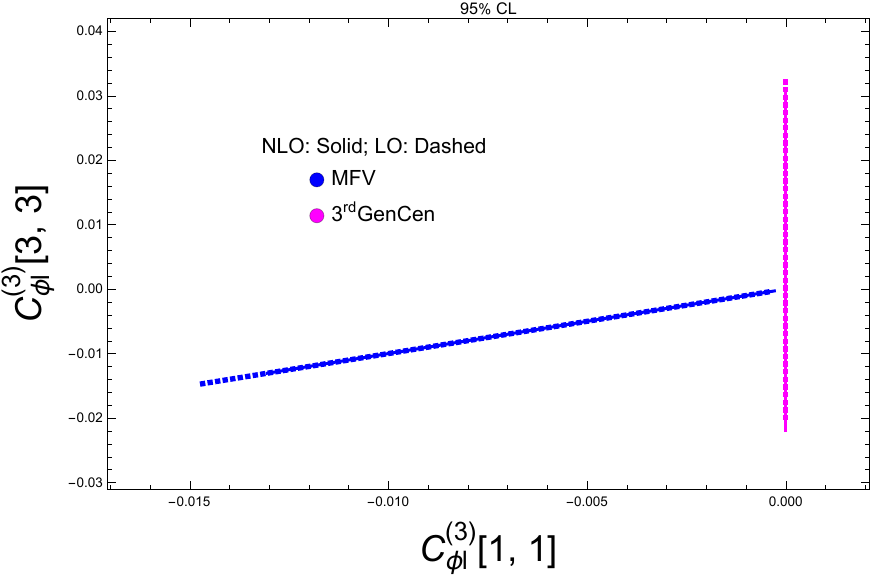}  \hskip .2in
\includegraphics[width=3in]{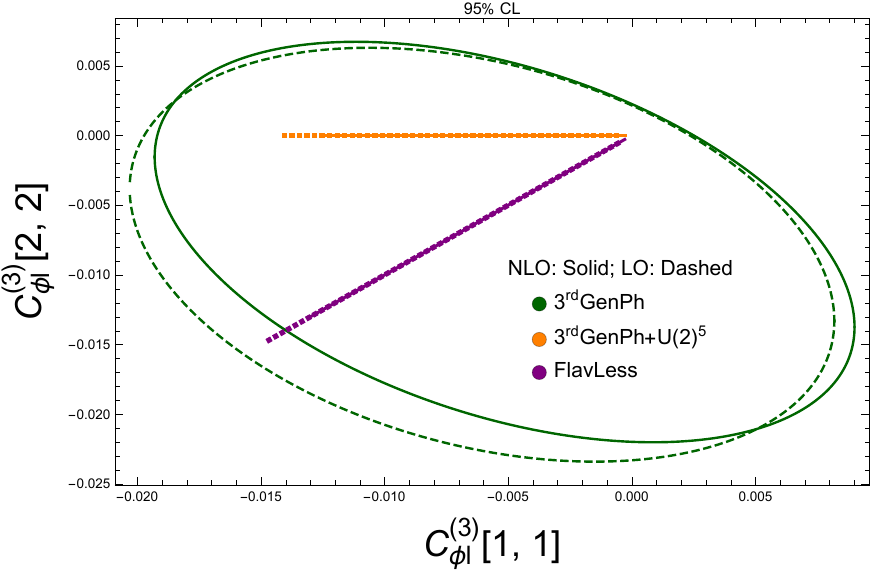} 
	\caption{95\% CL limits on $\C_{\phi l}^{(3)}[ij]$ under flavor assumptions described in the text. Results at LO are drawn with dashed lines, results at NLO are drawn with solid lines. {\bf On the left} we present $\C_{\phi l}^{(3)}[11]$ vs. $\C_{\phi l}^{(3)}[33]$ in the MFV (blue) and $3^{rd}$ generation centric (magenta) scenarios. In these scenarios $\C_{\phi l}^{(3)}[22]=\C_{\phi l}^{(3)}[11]$.  {\bf On the right} we present $\C_{\phi l}^{(3)}[11]$ vs. $\C_{\phi l}^{(3)}[22]$ in the $3^{rd}$ generation phobic (green), $3^{rd}$ generation phobic + $U(2)^5$ (orange) and flavorless (violet) scenarios. In the first two scenarios $\C_{\phi l}^{(3)}[33]=0$, while in the flavorless scenario $\C_{\phi l}^{(3)}[33]=\C_{\phi l}^{(3)}[22]=\C_{\phi l}^{(3)}[11]$. All other coefficients are set to 0.}
	\label{fig:fhl3}
\end{figure}

\end{document}